\documentclass[sigconf,10pt]{acmart}
\usepackage{float}
\usepackage{xcolor}
\usepackage{booktabs, siunitx}
\usepackage{graphicx}
\usepackage[acronyms,nonumberlist,nopostdot,nomain,nogroupskip,acronymlists={hidden}]{glossaries}
\usepackage{xspace}
\usepackage{pgfplots}
\pgfplotsset{compat=1.18}
\usepgfplotslibrary{statistics}
\usepackage{caption}
\usepackage{soul}
\usepackage{subcaption}
\usepackage{tikz}
\usepackage{multirow}
\usetikzlibrary{arrows.meta,positioning,fit,calc,shapes.geometric}
\usetikzlibrary{patterns}
\usepackage{booktabs}
\usepackage{tabularx}
\usepackage{array}
\usepackage{ragged2e}
\usepackage{enumitem}
\newglossary[algh]{hidden}{acrh}{acnh}{Hidden Acronyms}

\newcommand{\code}[1]{\texttt{\small #1}}
\newcolumntype{L}{>{\RaggedRight\arraybackslash}X}

\AtBeginDocument{%
  }

\copyrightyear{2026}
\acmYear{2026}
\setcopyright{cc}
\setcctype{by}
\acmConference[WiNTECH '26]{20th ACM Workshop on Wireless Network Testbeds, Experimental evaluation & Characterization }{October 26--30, 2026}{Austin, TX, USA}
\acmBooktitle{20th ACM Workshop on Wireless Network Testbeds, Experimental evaluation \& Characterization (WiNTECH '26), October 26--30, 2026, Austin, TX, USA}
\acmDOI{10.1145/3831662.3844165}
\acmISBN{979-8-4007-2879-2/2026/10}

\newif\ifexttikz
\exttikzfalse
\ifexttikz
\else
\usepackage{tikzpagenodes,etoolbox}
\usetikzlibrary{calc}
\usepackage[contents={}]{background}
\AddEverypageHook{%
\ifnumequal{\thepage}{1}{%
    \tikz[remember picture,overlay]{%
        \node[draw,
        minimum width=1.03\textwidth,
        text width=1.02\textwidth,
        font=\footnotesize
        ]
        at ($(current page header area) - (0, -8pt)$)
        {%
    This paper has been accepted for publication at the 20th ACM Workshop on
Wireless Network Testbeds, Experimental evaluation \& Characterization
(WiNTECH~'26), co-located with ACM MobiCom~2026, Austin, TX, USA,
October~30, 2026. \copyright~2026 Copyright held by the owner/author(s).
This work is licensed under a Creative Commons Attribution 4.0
International License (CC~BY~4.0). The definitive version is:
A.~Feraudo, R.~Shirkhani, S.~Maxenti, A.~Lacava, M.~Polese, and
T.~Melodia, ``Bringing dApps to OCUDU: An E3 Controller for Real-Time
Open RAN Intelligence,'' in \emph{Proc. 20th ACM Workshop on Wireless
Network Testbeds, Experimental evaluation \& Characterization
(WiNTECH~'26)}, ACM, New York, NY, USA, pp.~1--6.
        };
        \node[draw,
        minimum width=1.03\textwidth,
        text width=1.02\textwidth,
        font=\footnotesize
        ]
        at ($(current page footer area) - (0,20pt)$)
        {%
        © 2026 Copyright held by the owner/author(s). This work is licensed under a Creative Commons Attribution 4.0 International License. Published in the Proceedings of the 20th ACM Workshop on Wireless Network Testbeds, Experimental evaluation \& Characterization (WiNTECH '26), Austin, TX, USA, October 30, 2026. ACM, New York, NY, USA. \url{https://doi.org/10.1145/3831662.3844165}
        };
    }%
}{}
}
\fi

\begin{document}
\newacronym{3gpp}{3GPP}{3rd Generation Partnership Project}
\newacronym{4g}{4G}{4th generation}
\newacronym{5g}{5G}{5th generation}
\newacronym{5gc}{5GC}{5G Core}
\newacronym{6g}{6G}{6th generation}
\newacronym{adc}{ADC}{Analog to Digital Converter}
\newacronym{aerpaw}{AERPAW}{Aerial Experimentation and Research Platform for Advanced Wireless}
\newacronym{afosr}{AFOSR}{Air Force Office of Scientific Research}
\newacronym{afrl}{AFRL}{Air Force Research Laboratory}
\newacronym{ai}{AI}{Artificial Intelligence}
\newacronym{aimd}{AIMD}{Additive Increase Multiplicative Decrease}
\newacronym{am}{AM}{Acknowledged Mode}
\newacronym{amc}{AMC}{Adaptive Modulation and Coding}
\newacronym{amf}{AMF}{Access and Mobility Management Function}
\newacronym{aoa}{AoA}{Angle of Arrival}
\newacronym{aops}{AOPS}{Adaptive Order Prediction Scheduling}
\newacronym{ap}{AP}{Application Protocol}
\newacronym{api}{API}{Application Programming Interface}
\newacronym{apn}{APN}{Access Point Name}
\newacronym{aqm}{AQM}{Active Queue Management}
\newacronym{arc}{ARC}{Aerial RAN CoLab}
\newacronym{arl}{ARL}{Army Research Laboratory}
\newacronym{arpu}{ARPU}{Average Revenue per User}
\newacronym{asm}{ASM}{Advanced Sleep Mode}
\newacronym{asn1}{ASN.1}{Abstract Syntax Notation One}
\newacronym{ausf}{AUSF}{Authentication Server Function}
\newacronym{avc}{AVC}{Advanced Video Coding}
\newacronym{awgn}{AGWN}{Additive White Gaussian Noise}
\newacronym{balia}{BALIA}{Balanced Link Adaptation Algorithm}
\newacronym{bbu}{BBU}{Base Band Unit}
\newacronym{bdp}{BDP}{Bandwidth-Delay Product}
\newacronym{ber}{BER}{Bit Error Rate}
\newacronym{bf}{BF}{Beamforming}
\newacronym{bler}{BLER}{Block Error Rate}
\newacronym{brr}{BRR}{Bayesian Ridge Regressor}
\newacronym{bs}{BS}{Base Station}
\newacronym{bsr}{BSR}{Buffer Status Report}
\newacronym{bss}{BSS}{Business Support System}
\newacronym{bwp}{BWP}{Bandwidth Part}
\newacronym{ca}{CA}{Carrier Aggregation}
\newacronym{caas}{CaaS}{Connectivity-as-a-Service}
\newacronym{cae}{CAE}{Cognitive Analysis Engine}
\newacronym{cb}{CB}{Code Block}
\newacronym{cbrs}{CBRS}{Citizen Broadband Radio Service}
\newacronym{cc}{CC}{Congestion Control}
\newacronym{ccid}{CCID}{Congestion Control ID}
\newacronym{cco}{CC}{Carrier Component}
\newacronym{cd}{CD}{Continuous Delivery}
\newacronym{cdd}{CDD}{Cyclic Delay Diversity}
\newacronym{cdf}{CDF}{Cumulative Distribution Function}
\newacronym{cdn}{CDN}{Content Distribution Network}
\newacronym{cfr}{CFR}{Crest Factor Reduction}
\newacronym{cli}{CLI}{Command-line Interface}
\newacronym{cn}{CN}{Core Network}
\newacronym{cnn}{CNN}{Convolutional Neural Network}
\newacronym{codel}{CoDel}{Controlled Delay Management}
\newacronym{comac}{COMAC}{Converged Multi-Access and Core}
\newacronym{cord}{CORD}{Central Office Re-architected as a Datacenter}
\newacronym{cornet}{CORNET}{COgnitive Radio NETwork}
\newacronym{cosmos}{COSMOS}{Cloud Enhanced Open Software Defined Mobile Wireless Testbed for City-Scale Deployment}
\newacronym{cots}{COTS}{Commercial Off-the-Shelf}
\newacronym{cp}{CP}{Control Plane}
\newacronym{cpu}{CPU}{Central Processing Unit}
\newacronym{cqi}{CQI}{Channel Quality Information}
\newacronym{cr}{CR}{Cognitive Radio}
\newacronym{cran}{CRAN}{Cloud \gls{ran}}
\newacronym{crs}{CRS}{Cell Reference Signal}
\newacronym{csi}{CSI}{Channel State Information}
\newacronym{csirs}{CSI-RS}{Channel State Information - Reference Signal}
\newacronym{ct}{CT}{Continuous Testing}
\newacronym{cu}{CU}{Central Unit}
\newacronym{cuda}{CUDA}{Compute Unified Device Architecture}
\newacronym{cus}{CUS}{Control, User, Synchronization}
\newacronym{cve}{CVE}{Common Vulnerabilities and Exposure}
\newacronym{cyp}{CP}{Cyclic Prefix}
\newacronym{d2tcp}{D$^2$TCP}{Deadline-aware Data center TCP}
\newacronym{d3}{D$^3$}{Deadline-Driven Delivery}
\newacronym{dac}{DAC}{Digital to Analog Converter}
\newacronym{dag}{DAG}{Directed Acyclic Graph}
\newacronym{das}{DAS}{Distributed Antenna System}
\newacronym{dash}{DASH}{Dynamic Adaptive Streaming over HTTP}
\newacronym{dc}{DC}{Dual Connectivity}
\newacronym{dccp}{DCCP}{Datagram Congestion Control Protocol}
\newacronym{dce}{DCE}{Direct Code Execution}
\newacronym{dci}{DCI}{Downlink Control Information}
\newacronym{dctcp}{DCTCP}{Data Center TCP}
\newacronym{dfe}{DFE}{Digital Front-End}
\newacronym{dl}{DL}{Downlink}
\newacronym{dmr}{DMR}{Deadline Miss Ratio}
\newacronym{dmrs}{DMRS}{DeModulation Reference Signal}
\newacronym{dod}{DoD}{Department of Defense}
\newacronym{dos}{DoS}{Denial of Service}
\newacronym{dpd}{DPD}{Digital Pre-Distorsion}
\newacronym{dpu}{DPU}{Data Processing Unit}
\newacronym{drl}{DRL}{Deep Reinforcement Learning}
\newacronym{drlcc}{DRL-CC}{Deep Reinforcement Learning Congestion Control}
\newacronym{drs}{DRS}{Discovery Reference Signal}
\newacronym{dsp}{DSP}{Digital Signal Processing}
\newacronym{du}{DU}{Distributed Unit}
\newacronym{e2ap}{E2AP}{E2 Application Protocol}
\newacronym{e2e}{E2E}{end-to-end}
\newacronym{e2sm}{E2SM}{E2 Service Model}
\newacronym{e3ap}{E3AP}{E3 Application Protocol}
\newacronym{e3sm}{E3SM}{E3 Service Model}
\newacronym{earfcn}{EARFCN}{E-UTRA Absolute Radio Frequency Channel Number}
\newacronym{eaxcid}{eAxC\_ID}{extended Antenna-Carrier Identifier}
\newacronym{ecaas}{ECaaS}{Edge-Cloud-as-a-Service}
\newacronym{ecn}{ECN}{Explicit Congestion Notification}
\newacronym{ecpri}{eCPRI}{enhanced Common Public Radio Interface}
\newacronym{edf}{EDF}{Earliest Deadline First}
\newacronym{ei}{EI}{Enrichment Information}
\newacronym{embb}{eMBB}{Enhanced Mobile Broadband}
\newacronym{empower}{EMPOWER}{EMpowering transatlantic PlatfOrms for advanced WirEless Research}
\newacronym{enb}{eNB}{evolved Node Base}
\newacronym{endc}{EN-DC}{E-UTRAN-\gls{nr} \gls{dc}}
\newacronym{epc}{EPC}{Evolved Packet Core}
\newacronym{eps}{EPS}{Evolved Packet System}
\newacronym{es}{ES}{Edge Server}
\newacronym[firstplural=Estimated Times of Arrival (ETAs)]{eta}{ETA}{Estimated Time of Arrival}
\newacronym{etsi}{ETSI}{European Telecommunications Standards Institute}
\newacronym{eutran}{E-UTRAN}{Evolved Universal Terrestrial Access Network}
\newacronym{faas}{FaaS}{Function-as-a-Service}
\newacronym{fapi}{FAPI}{Functional Application Platform Interface}
\newacronym{fdd}{FDD}{Frequency Division Duplexing}
\newacronym{fdm}{FDM}{Frequency Division Multiplexing}
\newacronym{fdma}{FDMA}{Frequency Division Multiple Access}
\newacronym{fed4fire}{FED4FIRE+}{Federation 4 Future Internet Research and Experimentation Plus}
\newacronym{fft}{FFT}{Fast Fourier Transform}
\newacronym{fh}{FH}{Fronthaul}
\newacronym{fir}{FIR}{Finite Impulse Response}
\newacronym{fit}{FIT}{Future \acrlong{iot}}
\newacronym{fpga}{FPGA}{Field Programmable Gate Array}
\newacronym{fr}{FR}{Frequency Range}
\newacronym{fr2}{FR2}{Frequency Range 2}
\newacronym{fs}{FS}{Fast Switching}
\newacronym{fscc}{FSCC}{Flow Sharing Congestion Control}
\newacronym{ftp}{FTP}{File Transfer Protocol}
\newacronym{fw}{FW}{Flow Window}
\newacronym{gcpw}{GCPW}{Grounded Co-Planar Waveguide}
\newacronym{ge}{GE}{Gaussian Elimination}
\newacronym{gh}{GH}{Grace Hopper}
\newacronym{gmp}{GMP}{Generalized Memory Polynomial}
\newacronym{gnb}{gNB}{Next Generation Node Base}
\newacronym{gop}{GOP}{Group of Pictures}
\newacronym{gpio}{GPIO}{General Purpose Input/Output}
\newacronym{gpr}{GPR}{Gaussian Process Regressor}
\newacronym{gpu}{GPU}{Graphics Processing Unit}
\newacronym{gtp}{GTP}{GPRS Tunneling Protocol}
\newacronym{gtpc}{GTP-C}{GPRS Tunnelling Protocol Control Plane}
\newacronym{gtpu}{GTP-U}{GPRS Tunnelling Protocol User Plane}
\newacronym{gtpv2c}{GTPv2-C}{\gls{gtp} v2 - Control}
\newacronym{gw}{GW}{Gateway}
\newacronym{harq}{HARQ}{Hybrid Automatic Repeat reQuest}
\newacronym{hbom}{HBOM}{Hardware Bill of Materials}
\newacronym{hetnet}{HetNet}{Heterogeneous Network}
\newacronym{hh}{HH}{Hard Handover}
\newacronym{hol}{HOL}{Head-of-Line}
\newacronym{hqf}{HQF}{Highest-quality-first}
\newacronym{hss}{HSS}{Home Subscription Server}
\newacronym{http}{HTTP}{HyperText Transfer Protocol}
\newacronym{ia}{IA}{Initial Access}
\newacronym{iab}{IAB}{Integrated Access and Backhaul}
\newacronym{ic}{IC}{Incident Command}
\newacronym{ietf}{IETF}{Internet Engineering Task Force}
\newacronym{if}{IF}{Intermediate Frequency}
\newacronym{imsi}{IMSI}{International Mobile Subscriber Identity}
\newacronym{imt}{IMT}{International Mobile Telecommunication}
\newacronym{iot}{IoT}{Internet of Things}
\newacronym{ip}{IP}{Internet Protocol}
\newacronym{ipc}{IPC}{Inter-Process Communication}
\newacronym{isac}{ISAC}{Integrated Sensing and Communication}
\newacronym{itu}{ITU}{International Telecommunication Union}
\newacronym{kpi}{KPI}{Key Performance Indicator}
\newacronym{kpm}{KPM}{Key Performance Measurement}
\newacronym{kvm}{KVM}{Kernel-based Virtual Machine}
\newacronym{laa}{LAA}{Licensed-Assisted Access}
\newacronym{lcm}{LCM}{Lifecycle Management}
\newacronym{lna}{LNA}{Low-Noise Amplifier}
\newacronym{lo}{LO}{Local Oscillator}
\newacronym{los}{LOS}{Line-of-Sight}
\newacronym{lsm}{LSM}{Link-to-System Mapping}
\newacronym{lstm}{LSTM}{Long Short Term Memory}
\newacronym{lte}{LTE}{Long Term Evolution}
\newacronym{lxc}{LXC}{Linux Container}
\newacronym{m2m}{M2M}{Machine to Machine}
\newacronym{mac}{MAC}{Medium Access Control}
\newacronym{macsec}{MACsec}{Media Access Control Security}
\newacronym{manet}{MANET}{Mobile Ad Hoc Network}
\newacronym{mano}{MANO}{Management and Orchestration}
\newacronym{mc}{MC}{Multi-Connectivity}
\newacronym{mcc}{MCC}{Mobile Cloud Computing}
\newacronym{mchem}{MCHEM}{Massive Channel Emulator}
\newacronym{mcs}{MCS}{Modulation and Coding Scheme}
\newacronym{mec}{MEC}{Multi-access Edge Computing}
\newacronym{mec2}{MEC}{Mobile Edge Cloud}
\newacronym{mfc}{MFC}{Mobile Fog Computing}
\newacronym{mgen}{MGEN}{Multi-Generator}
\newacronym{mi}{MI}{Mutual Information}
\newacronym{mib}{MIB}{Master Information Block}
\newacronym{miesm}{MIESM}{Mutual Information Based Effective SINR}
\newacronym{mimo}{MIMO}{Multiple Input, Multiple Output}
\newacronym{ml}{ML}{Machine Learning}
\newacronym{mlr}{MLR}{Maximum-local-rate}
\newacronym[plural=\gls{mme}s,firstplural=Mobility Management Entities (MMEs)]{mme}{MME}{Mobility Management Entity}
\newacronym{mmtc}{mMTC}{Massive Machine-Type Communications}
\newacronym{mmwave}{mmWave}{millimeter wave}
\newacronym{mno}{MNO}{Mobile Network Operator}
\newacronym{mns}{MnS}{Management Services}
\newacronym{mpdccp}{MP-DCCP}{Multipath Datagram Congestion Control Protocol}
\newacronym{mptcp}{MPTCP}{Multipath TCP}
\newacronym{mr}{MR}{Maximum Rate}
\newacronym{mrdc}{MR-DC}{Multi \gls{rat} \gls{dc}}
\newacronym{mrl}{MRL}{Manufacturing Readiness Level}
\newacronym{mse}{MSE}{Mean Square Error}
\newacronym{mss}{MSS}{Maximum Segment Size}
\newacronym{mt}{MT}{Mobile Termination}
\newacronym{mtc}{MTC}{Machine-type Communications}
\newacronym{mtd}{MTD}{Machine-Type Device}
\newacronym{mtu}{MTU}{Maximum Transmission Unit}
\newacronym{mumimo}{MU-MIMO}{Multi-user \gls{mimo}}
\newacronym{mvno}{MVNO}{Mobile Virtual Network Operator}
\newacronym{nalu}{NALU}{Network Abstraction Layer Unit}
\newacronym{nas}{NAS}{Network Attached Storage}
\newacronym{nat}{NAT}{Network Address Translation}
\newacronym{nbiot}{NB-IoT}{Narrow Band IoT}
\newacronym{nco}{NCO}{Numerically Controlled Oscillator}
\newacronym{nfv}{NFV}{Network Function Virtualization}
\newacronym{nfvi}{NFVI}{Network Function Virtualization Infrastructure}
\newacronym{ngrg}{nGRG}{next Generation Research Group}
\newacronym{ni}{NI}{Network Interfaces}
\newacronym{nic}{NIC}{Network Interface Card}
\newacronym{nlos}{NLOS}{Non-Line-of-Sight}
\newacronym{now}{NOW}{Non Overlapping Window}
\newacronym[type=hidden]{nr}{NR}{New Radio}
\newacronym{nrf}{NRF}{Network Repository Function}
\newacronym{nrric}{Near-RT RIC}{Near-Real-Time RAN Intelligent Controller}
\newacronym{nsa}{NSA}{Non Stand Alone}
\newacronym{nse}{NSE}{Network Slicing Engine}
\newacronym{nsin}{NSIN}{National Security Innovation Network}
\newacronym{nsm}{NSM}{Network Service Mesh}
\newacronym{nssf}{NSSF}{Network Slice Selection Function}
\newacronym{nyu}{NYU}{New York University}
\newacronym{o2i}{O2I}{Outdoor to Indoor}
\newacronym{oai}{OAI}{OpenAirInterface}
\newacronym{oaicn}{OAI-CN}{\gls{oai} \acrlong{cn}}
\newacronym{oairan}{OAI-RAN}{\acrlong{oai} \acrlong{ran}}
\newacronym{oam}{OAM}{Operations, Administration and Maintenance}
\newacronym{oci}{OCI}{Open Container Initiative}
\newacronym{oem}{OEM}{Original Equipment Manufacturer}
\newacronym{ofdm}{OFDM}{Orthogonal Frequency Division Multiplexing}
\newacronym{olia}{OLIA}{Opportunistic Linked Increase Algorithm}
\newacronym{omec}{OMEC}{Open Mobile Evolved Core}
\newacronym{onap}{ONAP}{Open Network Automation Platform}
\newacronym{onf}{ONF}{Open Networking Foundation}
\newacronym{onos}{ONOS}{Open Networking Operating System}
\newacronym{onr}{ONR}{Office of Naval Research}
\newacronym{oom}{OOM}{\gls{onap} Operations Manager}
\newacronym{opex}{OPEX}{operational expenses}
\newacronym{opnfv}{OPNFV}{Open Platform for \gls{nfv}}
\newacronym[type=hidden]{oran}{O-RAN}{Open \gls{ran}}
\newacronym{orbit}{ORBIT}{Open-Access Research Testbed for Next-Generation Wireless Networks}
\newacronym{os}{OS}{Operating System}
\newacronym{osc}{OSC}{O-RAN Software Community}
\newacronym{osm}{OSM}{Open Source \gls{nfv} Management and Orchestration}
\newacronym{oss}{OSS}{Operations Support System}
\newacronym{ota}{OTA}{Over-The-Air}
\newacronym{json}{JSON}{JavaScript Object Notation}
\newacronym{pa}{PA}{Position-aware}
\newacronym{papr}{PAPR}{Peak-to-Average Power Ratio}
\newacronym{pase}{PASE}{Prioritization, Arbitration, and Self-adjusting Endpoints}
\newacronym{pawr}{PAWR}{Platforms for Advanced Wireless Research}
\newacronym{pbch}{PBCH}{Physical Broadcast Channel}
\newacronym{pcb}{PCB}{Printed Circuit Board}
\newacronym{pcef}{PCEF}{Policy and Charging Enforcement Function}
\newacronym{pcfich}{PCFICH}{Physical Control Format Indicator Channel}
\newacronym{pci}{PCIe}{Peripheral Component Interconnect Express}
\newacronym{pcrf}{PCRF}{Policy and Charging Rules Function}
\newacronym{pdcch}{PDCCH}{Physical Downlink Control Channel}
\newacronym{pdcp}{PDCP}{Packet Data Convergence Protocol}
\newacronym{pdf}{PDF}{Probability Density Function}
\newacronym{pdsch}{PDSCH}{Physical Downlink Shared Channel}
\newacronym{pdu}{PDU}{Packet Data Unit}
\newacronym{pf}{PF}{Proportional Fair}
\newacronym{pgw}{PGW}{Packet Gateway}
\newacronym{phich}{PHICH}{Physical Hybrid ARQ Indicator Channel}
\newacronym{phy}{PHY}{Physical}
\newacronym{plfs}{PLFS}{Physical Layer Frequency Signals}
\newacronym{pll}{PLL}{Phased-Locked Loop}
\newacronym{pmch}{PMCH}{Physical Multicast Channel}
\newacronym{pmi}{PMI}{Precoding Matrix Indicators}
\newacronym{pnf}{PNF}{Physical Network Function}
\newacronym{powder}{POWDER}{Platform for Open Wireless Data-driven Experimental Research}
\newacronym{ppo}{PPO}{Proximal Policy Optimization}
\newacronym{ppp}{PPP}{Poisson Point Process}
\newacronym{prach}{PRACH}{Physical Random Access Channel}
\newacronym{prb}{PRB}{Physical Resource Block}
\newacronym{pse}{PSE}{Performance Specialized Engine}
\newacronym{psnr}{PSNR}{Peak Signal to Noise Ratio}
\newacronym{pss}{PSS}{Primary Synchronization Signal}
\newacronym{pt}{PT}{Plain Text}
\newacronym{ptp}{PTP}{Precision Time Protocol}
\newacronym{pucch}{PUCCH}{Physical Uplink Control Channel}
\newacronym{pusch}{PUSCH}{Physical Uplink Shared Channel}
\newacronym{qam}{QAM}{Quadrature Amplitude Modulation}
\newacronym{qci}{QCI}{\gls{qos} Class Identifier}
\newacronym{qoe}{QoE}{Quality of Experience}
\newacronym{qos}{QoS}{Quality of Service}
\newacronym{qsfp}{QSFP}{quad small form factor pluggable}
\newacronym{quic}{QUIC}{Quick UDP Internet Connections}
\newacronym{rach}{RACH}{Random Access Channel}
\newacronym{ran}{RAN}{Radio Access Network}
\newacronym{rapl}{RAPL}{Running Average Power Limit}
\newacronym[firstplural=Radio Access Technologies (RATs)]{rat}{RAT}{Radio Access Technology}
\newacronym{rbg}{RBG}{Resource Block Group}
\newacronym{rc}{RC}{RAN Control}
\newacronym{rcn}{RCN}{Research Coordination Network}
\newacronym{rec}{REC}{Radio Edge Cloud}
\newacronym{red}{RED}{Random Early Detection}
\newacronym{renew}{RENEW}{Reconfigurable Eco-system for Next-generation End-to-end Wireless}
\newacronym{rf}{RF}{Radio Frequency}
\newacronym{rfc}{RFC}{Request for Comments}
\newacronym{rfr}{RFR}{Random Forest Regressor}
\newacronym{rfsoc}{RFSoC}{Radio Frequency System-on-Chip}
\newacronym{ric}{RIC}{RAN Intelligent Controller}
\newacronym{rlc}{RLC}{Radio Link Control}
\newacronym{rlf}{RLF}{Radio Link Failure}
\newacronym{rlnc}{RLNC}{Random Linear Network Coding}
\newacronym{rmr}{RMR}{RIC Message Router}
\newacronym{rmse}{RMSE}{Root Mean Squared Error}
\newacronym{rnis}{RNIS}{Radio Network Information Service}
\newacronym{rnti}{RNTI}{Radio Network Temporary Identifier}
\newacronym{rr}{RR}{Round Robin}
\newacronym{rrc}{RRC}{Radio Resource Control}
\newacronym{rrm}{RRM}{Radio Resource Management}
\newacronym{rru}{RRU}{Remote Radio Unit}
\newacronym{rs}{RS}{Remote Server}
\newacronym{rsrp}{RSRP}{Reference Signal Received Power}
\newacronym{rsrq}{RSRQ}{Reference Signal Received Quality}
\newacronym{rss}{RSS}{Received Signal Strength}
\newacronym{rssi}{RSSI}{Received Signal Strength Indicator}
\newacronym{rt}{RT}{Real-Time}
\newacronym{rtt}{RTT}{Round Trip Time}
\newacronym{ru}{RU}{Radio Unit}
\newacronym{rw}{RW}{Receive Window}
\newacronym{rx}{RX}{Receiver}
\newacronym{s1ap}{S1AP}{S1 Application Protocol}
\newacronym{sa}{SA}{standalone}
\newacronym{sack}{SACK}{Selective Acknowledgment}
\newacronym{sap}{SAP}{Service Access Point}
\newacronym{sbom}{SBOM}{Software Bill of Materials}
\newacronym{sc2}{SC2}{Spectrum Collaboration Challenge}
\newacronym{scef}{SCEF}{Service Capability Exposure Function}
\newacronym{sch}{SCH}{Secondary Cell Handover}
\newacronym{scoot}{SCOOT}{Split Cycle Offset Optimization Technique}
\newacronym{sctp}{SCTP}{Stream Control Transmission Protocol}
\newacronym{sdap}{SDAP}{Service Data Adaptation Protocol}
\newacronym{sdk}{SDK}{Software Development Kit}
\newacronym{sdl}{SDL}{Shared Data Layer}
\newacronym{sdm}{SDM}{Space Division Multiplexing}
\newacronym{sdma}{SDMA}{Spatial Division Multiple Access}
\newacronym{sdn}{SDN}{Software-defined Networking}
\newacronym{sdr}{SDR}{Software-defined Radio}
\newacronym{seba}{SEBA}{SDN-Enabled Broadband Access}
\newacronym{sgsn}{SGSN}{Serving GPRS Support Node}
\newacronym{sgw}{SGW}{Service Gateway}
\newacronym{si}{SI}{Study Item}
\newacronym{sib}{SIB}{Secondary Information Block}
\newacronym{simd}{SIMD}{Single Instruction/Multiple Data}
\newacronym{sinr}{SINR}{Signal to Interference plus Noise Ratio}
\newacronym{sip}{SIP}{Session Initiation Protocol}
\newacronym{siso}{SISO}{Single Input, Single Output}
\newacronym{sla}{SLA}{Service Level Agreement}
\newacronym{sm}{SM}{Service Model}
\newacronym{sme}{SME}{Small-Medium Enterprise}
\newacronym{smf}{SMF}{Session Management Function}
\newacronym{smo}{SMO}{Service Management and Orchestration}
\newacronym{sms}{SMS}{Short Message Service}
\newacronym{smsgmsc}{SMS-GMSC}{\gls{sms}-Gateway}
\newacronym{snr}{SNR}{Signal-to-Noise-Ratio}
\newacronym{som}{SOM}{System-on-Module}
\newacronym{son}{SON}{Self-Organizing Network}
\newacronym{spi}{SPI}{Serial Peripheral Interface}
\newacronym{spsc}{SPSC}{single-producer/single-consumer}
\newacronym{sptcp}{SPTCP}{Single Path TCP}
\newacronym{srb}{SRB}{Service Radio Bearer}
\newacronym{srfa}{SRFA}{Special Research Focus Area}
\newacronym{srn}{SRN}{Standard Radio Node}
\newacronym{srs}{SRS}{Sounding Reference Signal}
\newacronym{ss}{SS}{Synchronization Signal}
\newacronym{sss}{SSS}{Secondary Synchronization Signal}
\newacronym{st}{ST}{Spanning Tree}
\newacronym{svc}{SVC}{Scalable Video Coding}
\newacronym{tb}{TB}{Transport Block}
\newacronym{tbs}{TBS}{Transport Block Size}
\newacronym{tcp}{TCP}{Transmission Control Protocol}
\newacronym{tdd}{TDD}{Time Division Duplexing}
\newacronym{tdm}{TDM}{Time Division Multiplexing}
\newacronym{tdma}{TDMA}{Time Division Multiple Access}
\newacronym{tfl}{TfL}{Transport for London}
\newacronym{tfrc}{TFRC}{TCP-Friendly Rate Control}
\newacronym{tft}{TFT}{Traffic Flow Template}
\newacronym{tgen}{TGEN}{Traffic Generator}
\newacronym{tip}{TIP}{Telecom Infra Project}
\newacronym{tm}{TM}{Transparent Mode}
\newacronym{to}{TO}{Telco Operator}
\newacronym{tr}{TR}{Technical Report}
\newacronym{trl}{TRL}{Technology Readiness Level}
\newacronym{trp}{TRP}{Transmitter Receiver Pair}
\newacronym{ts}{TS}{Technical Specification}
\newacronym{tti}{TTI}{Transmission Time Interval}
\newacronym{ttt}{TTT}{Time-to-Trigger}
\newacronym{tx}{TX}{Transmitter}
\newacronym{txb}{TXB}{Transmit Beam}
\newacronym{uas}{UAS}{Unmanned Aerial System}
\newacronym{uav}{UAV}{Unmanned Aerial Vehicle}
\newacronym{udm}{UDM}{Unified Data Management}
\newacronym{udp}{UDP}{User Datagram Protocol}
\newacronym{udr}{UDR}{Unified Data Repository}
\newacronym{ue}{UE}{User Equipment}
\newacronym{uhd}{UHD}{\gls{usrp} Hardware Driver}
\newacronym{ul}{UL}{Uplink}
\newacronym{ulpi}{ULPI}{Uplink Performance Improvement}
\newacronym{um}{UM}{Unacknowledged Mode}
\newacronym{uml}{UML}{Unified Modeling Language}
\newacronym{up}{UP}{User Plane}
\newacronym{upa}{UPA}{Uniform Planar Array}
\newacronym{upf}{UPF}{User Plane Function}
\newacronym{urllc}{URLLC}{Ultra Reliable and Low Latency Communications}
\newacronym{usa}{U.S.}{United States}
\newacronym{usim}{USIM}{Universal Subscriber Identity Module}
\newacronym{usrp}{USRP}{Universal Software Radio Peripheral}
\newacronym{utc}{UTC}{Urban Traffic Control}
\newacronym{ves}{VES}{\gls{vnf} Event Stream}
\newacronym{vex}{VEX}{Vulnerability Exploitability eXchange}
\newacronym{vim}{VIM}{Virtualization Infrastructure Manager}
\newacronym{vm}{VM}{Virtual Machine}
\newacronym{vnf}{VNF}{Virtual Network Function}
\newacronym{volte}{VoLTE}{Voice over \gls{lte}}
\newacronym{voltha}{VOLTHA}{Virtual OLT HArdware Abstraction}
\newacronym{vr}{VR}{Virtual Reality}
\newacronym{vran}{vRAN}{Virtualized \gls{ran}}
\newacronym{vss}{VSS}{Video Streaming Server}
\newacronym{wbf}{WBF}{Wired Bias Function}
\newacronym{wf}{WF}{Waterfilling}
\newacronym{wg}{WG}{Working Group}
\newacronym{wlan}{WLAN}{Wireless Local Area Network}

\glsunset{ran}

\newcommand\note[2]{\color{#1}\bf #2}
\newcommand\af[1]{{\note{blue}{angelo: #1}}}
\newcommand\sm[1]{{\note{orange}{stefano: #1}}}
\newcommand\rs[1]{{\note{green}{ravis: #1}}}
\newcommand\mpo[1]{{\note{purple}{michele: #1}}}
\newcommand\tm[1]{{\note{red}{tommaso: #1}}}

\newcommand*{\phyl}{PHY-low\xspace}
\newcommand*{\pci}{\gls{pci}\xspace}
\newcommand*{\aiml}{\gls{ai}/\gls{ml}\xspace}
\newcommand{\ric}{\gls{ric}\xspace}
\newcommand{\rics}{\glspl{ric}\xspace}
\newcommand{\nearrt}{Near-\gls{rt}\xspace}
\newcommand{\nonrt}{Non-\gls{rt}\xspace}
\newcommand{\eax}{\gls{eaxcid}\xspace}
\newcommand{\ran}{\gls{ran}\xspace}

\definecolor{codegray}{rgb}{0.25,0.25,0.25}
\definecolor{codepurple}{rgb}{0.58,0,0.82}

\title{Bringing dApps to OCUDU: An E3 Controller for Real-Time Open RAN Intelligence}

\author[A. Feraudo, R. Shirkhani, S. Maxenti, A. Lacava, M. Polese, T. Melodia]{Angelo Feraudo,
Ravis Shirkhani,
Stefano Maxenti,
Andrea Lacava,
Michele Polese, Tommaso Melodia
}
\affiliation{%
    \institution{Institute for Intelligent Networked Systems, Northeastern University, Boston, MA, U.S.A.}
    \city{}
  \state{}
  \country{}
  }

\renewcommand{\shortauthors}{Feraudo et al.}

\begin{abstract}
Real-time control loops in Open RAN are increasingly co-located with the gNB, where dApps, i.e., programmable applications with sub-millisecond access to PHY- and MAC-layer signals, enable latency-critical use cases such as spectrum sharing, channel-aware scheduling, and integrated sensing. What turns dApps from a per-vendor mechanism into a portable component of the emerging AI-RAN ecosystem is the \emph{E3 interface}: this interface defines how a dApp subscribes to RAN telemetry, receives indications, and issues control actions back to the RAN. To date, E3 has been implemented on OpenAirInterface and on NVIDIA Aerial; OCUDU, i.e., the Linux Foundation's open-source CU/DU project, has lacked a comparable E3 path. We close that gap with an open-source \texttt{E3Controller} for OCUDU, designed as a sidecar daemon so the E3 protocol stack, service-model logic, and dApp interactions live entirely outside RAN core processing, which is touched only through well-defined instrumentation hooks. We validate the controller with two reference service models, \emph{Spectrum} and \emph{L1}, that deliver O-RAN fronthaul I/Q samples to a spectrum-sensing dApp over a Foxconn RPQN~4800 radio unit. Our evaluation shows that the E3 path imposes no measurable throughput penalty on the OCUDU RAN, demonstrating that a real-time dApp/RAN interface is feasible on production-grade open RAN software.
\end{abstract}



\maketitle

\section{Introduction}

The Open RAN architecture, as defined by the O-RAN ALLIANCE, promotes programmable, multi-vendor cellular networks
. Its programmability is anchored on two \glspl{ric}: the \nearrt~\ric hosts \emph{xApps}, which apply control loops on the order of 10-1000~ms over the E2 interface, while the \nonrt~\ric hosts \emph{rApps} for longer-horizon policies. A growing set of use cases \cite{airan2024whitepaper,3gpp38765,poleseunderstanding2023,dorodapps2022}, however, sits below this latency floor: spectrum sensing, AI-driven channel estimation, beam tracking, integrated sensing, and per-slot scheduling all need access to PHY- and MAC-layer signals at the slot or symbol granularity. The E2 interface cannot carry these signals, either because they are too data-intensive or because operators consider them too sensitive to expose outside the \ran node.

To close this gap, \emph{dApps} were introduced as an extension of the O-RAN architecture~\cite{dorodapps2022,oran2024dapps}: lightweight applications co-located with the \ran node that read internal data directly and react in the hundreds-of-microseconds regime. What distinguishes a dApp from an ad-hoc \ran extension, however, is the E3 interface, a well-defined protocol for interacting with \ran components. This interface defines how a dApp subscribes to a class of telemetry, receives indications carrying the requested data, and issues control actions back to the \ran. By cleanly separating \emph{what} dApps exchange from \emph{how} a given \ran exposes it, E3 lets the same dApp run across stacks and the same \ran serve different dApps, turning dApps from per-vendor add-ons into an open layer of the emerging AI-\ran ecosystem. 

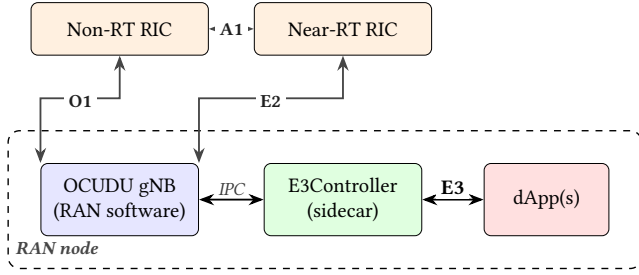
\begin{figure}[t]
\centering
\resizebox{\linewidth}{!}{
\begin{tikzpicture}[
    >=Stealth,
    box/.style={
        draw, rounded corners=3pt,
        minimum width=2.4cm, minimum height=1.1cm,
        align=center, font=\small,
    },
    ric/.style={box, fill=orange!12, minimum height=0.8cm, minimum width=2.7cm},
    gnb/.style={box, fill=blue!10},
    sidecar/.style={box, fill=green!12},
    dapp/.style={box, fill=red!12, minimum width=1.9cm},
    host/.style={
        draw, dashed, rounded corners=5pt,
        inner sep=14pt, line width=0.6pt,
    },
    biface/.style={<->, thick},
    e3iface/.style={<->, thick},
    oif/.style={<->, thick, gray!55!black},
    biflabel/.style={
        font=\footnotesize\itshape, text=gray!45!black,
        midway, above, inner sep=1.5pt
    },
    e3lbl/.style={
        font=\small\bfseries, midway, above,
        fill=white, inner sep=2pt, text=black
    },
    oiflbl/.style={
        font=\footnotesize\bfseries, text=gray!25!black,
        fill=white, inner sep=2pt
    },
    hostlabel/.style={
        font=\footnotesize\bfseries\itshape, text=gray!55!black,
        anchor=south west
    },
]
    \node[gnb]     (gnb)  at (0,   0) {OCUDU gNB\\(\ran software)};
    \node[sidecar] (ec)   at (3.4, 0) {E3Controller\\(sidecar)};
    \node[dapp]    (dapp) at (6.5, 0) {dApp(s)};

    \node[host, fit=(gnb)(ec)(dapp),
          label={[hostlabel, anchor=south west, yshift=3pt]south west: \ran node}] (host) {};

    \node[ric] (nrt)  at (0,   2.6) {Non-RT RIC};
    \node[ric] (nert) at (3.4, 2.6) {Near-RT RIC};

    \draw[biface] (gnb.east) -- (ec.west)   node[biflabel] {IPC};
    \draw[e3iface] (ec.east) -- (dapp.west) node[e3lbl] {E3};

    \draw[oif] (nrt.east) -- (nert.west) node[midway, oiflbl] {A1};
    \draw[oif] (nrt.south) -- ++(0, -0.65) -| (gnb.north west);
    \node[oiflbl] at (-0.6, 1.5) {O1};
    \draw[oif] (nert.south) -- ++(0, -0.65) -| (gnb.north east);
    \node[oiflbl] at (2.3, 1.5) {E2};
\end{tikzpicture}}
\caption{Deployment overview of the \texttt{E3Controller}.}
\label{fig:control-tiers}
\Description{Layered deployment diagram of the Open RAN control tiers. At the top, the Non-RT RIC hosts rApps for longer-horizon policies; in the middle, the Near-RT RIC hosts xApps that apply control loops over the E2 interface; at the bottom, dApps are co-located with the OCUDU gNB on the edge host. The \texttt{E3Controller} runs as a sidecar process on the same edge host as the gNB and exposes the E3 interface to one or more co-located dApps.}
\vspace{-.5cm}
\end{figure}

To date, E3 has been implemented on OpenAirInterface, where the reference dApp framework was first introduced~\cite{lacavadapps2025}, and on the NVIDIA Aerial GPU stack~\cite{gpu_isac}. OCUDU~\cite{ocudu2026} (old srsRAN), formerly srsRAN and increasingly common target for production-grade Open \ran trials, has, by contrast, lacked a comparable E3 path. Existing efforts that bring \ran-side telemetry to srsRAN/OCUDU, including the jbpf-enabled srsRAN fork from Microsoft~\cite{foukas2025srsranjbpf}, deliver low-latency instrumentation but do not expose it through a well-defined, dApp-facing interface.

This paper closes that gap by presenting \texttt{E3Controller}, an open-source framework that brings the E3 interface to OCUDU through a sidecar architecture. Figure~\ref{fig:control-tiers} provides an overview of the deployment: the \texttt{E3Controller} runs as a sidecar process on the same edge host as the OCUDU \gls{gnb}, and \emph{E3} is the interface it exposes to one or more co-located dApps. The E3 protocol stack, service-model logic, dApp interactions, and dApps all execute outside the OCUDU process, requiring lightweight instrumentation hooks in the \ran software stacks. In our deployment, we added two hooks to the OCUDU \gls{gnb}: one in the \gls{fh} receive path and one at the upper \gls{phy} level. These hooks serve as attachment points for jbpf~\cite{jbpf} codelets, leveraging Microsoft's open-source eBPF framework for instrumenting network functions, and compile to no-ops when the feature is disabled. To interact with dApps, the \texttt{E3Controller} uses the open-source \texttt{libE3} agent~\cite{libe3} for \gls{e3ap} session management and communication. We demonstrate the framework using two reference \glspl{e3sm}, \emph{Spectrum} and \emph{L1}. The \emph{Spectrum} service model supports both information reporting and E3 control actions, whereas the \emph{L1} service model streams O-RAN fronthaul I/Q samples. In the current \texttt{E3Controller} implementation, both service models support only the reporting interface, while E3 control actions remain under development. The experimental evaluation focuses on the \emph{L1} service model, as it represents the most demanding data path currently supported by the E3Controller, delivering high-rate fronthaul I/Q samples through shared memory. The experiments are performed on the X5G testbed~\cite{villa2025x5g}, using a commercial Foxconn RPQN-4800 radio unit, and a Samsung S23 \gls{ue}. The results show that the proposed E3 data path introduces no measurable throughput overhead to the OCUDU \ran.

\textbf{Contributions.} This paper makes the following contributions:
\begin{enumerate}[leftmargin=1.4em]
    \item The first end-to-end integration of the E3 interface with OCUDU, designed around a minimal-touch, sidecar architecture.
    \item An \texttt{E3Controller} that decouples the E3 interface from the underlying \ran software stack, enabling its adoption across different implementations. It features (i) dynamic codelet loading on dApp subscription via jbpf's \gls{lcm} \gls{ipc}, (ii) a three-stage pipeline with each stage pinnable to a dedicated core to bound tail latency, and (iii) a stream-multiplexed dispatcher with a zero-copy buffer hand-off that fans one codelet's data out to many dApps and serves many data types to one dApp through per-\ran-function subscriptions.
    \item An extension of the OCUDU \ran with two jbpf hooks that compile to no-ops when jbpf is disabled and require no changes to the OCUDU core. We pair them with two reference service models, \emph{Spectrum} (per-\gls{fh}-section, with sample decompression) and \emph{L1} (per-slot, all RX antennas), and validate the full path on a commercial O-RAN~7.2x \gls{ru}.
\end{enumerate}





\begin{figure*}[t]
\centering
\resizebox{\linewidth}{!}{\begin{tikzpicture}[
    box/.style={draw, rounded corners=3pt, minimum width=2.6cm, minimum height=1.05cm, align=center, font=\small},
    bigbox/.style={draw, dashed, rounded corners=5pt, inner sep=14pt},
    arrow/.style={-{Stealth[length=6pt]}, thick},
    alabel/.style={font=\small\itshape, text=gray!70!black, align=center},
    blabel/.style={font=\small\bfseries},
]
    \node[box, fill=blue!10]   (ran)   at ( 0.0, 0) {OCUDU\\(jbpf hook)};
    \node[box, fill=orange!15] (cod)   at ( 3.3, 0) {jbpf codelet};
    \node[bigbox, fit=(ran)(cod),
          label={[anchor=south west, yshift=4pt, blabel]north west:OCUDU gNB Process}] (gnb) {};

    \node[box, fill=yellow!20] (shm)   at ( 6.6, 0) {jbpf shared\\memory};

    \node[box, fill=green!10]  (disp)  at ( 9.9, 0) {JbpfDispatcher\\(poll thread)};
    \node[box, fill=green!10]  (sm)    at (13.2, 0) {Service Model\\(worker thread)};
    \node[box, fill=green!10]  (agent) at (16.5, 0) {libe3 publisher\\(APER / JSON)};
    \node[bigbox, fit=(disp)(sm)(agent),
          label={[anchor=south west, yshift=4pt, blabel]north west:E3Controller Daemon}] (e3c) {};

    \node[box, fill=red!10]    (dapp)  at (19.8, 0) {dApp};

    \draw[arrow] (ran.east)   -- (cod.west)   node[midway, above, alabel] {hook\\fires};
    \draw[arrow] (cod.east)   -- (shm.west);
    \draw[arrow] (shm.east)   -- (disp.west)  node[midway, above, alabel] {poll};
    \draw[arrow] (disp.east)  -- (sm.west)    node[midway, above, alabel] {};
    \draw[arrow] (sm.east)    -- (agent.west) node[midway, above, alabel] {};
    \draw[arrow] (agent.east) -- (dapp.west)
        node[midway, above, alabel] {E3\\Ind};

    \draw[arrow, dashed, blue!60!black] (dapp.south) -- ++(0,-1) -| (sm.south)
        node[pos=0.25, below, alabel] {Control Action};
    \draw[arrow, dashed, orange!70!black] (disp.south) -- ++(0,-1) -| ($(cod.south)+(0.25,0)$)
        node[pos=0.25, below, alabel] {LCM load (dynamic)};
\end{tikzpicture}}
\caption{End-to-end data flow from the OCUDU gNB to a subscribed dApp through the \texttt{E3Controller} sidecar.}
\label{fig:flow}
\Description{Block diagram tracing the end-to-end data path from the OCUDU gNB, on the left, to a subscribed dApp on the right. Inside the gNB, jbpf hooks invoke codelets that extract E3-relevant data and write it to a jbpf output channel identified by a unique stream_id. The data crosses a POSIX shared-memory IPC boundary into the \texttt{E3Controller} sidecar, where a three-stage pipeline processes it: the poll stage drains the shared-memory channel, the worker stage applies a service-model-specific transform and builds an Indication message, and the publisher stage encodes it in \gls{asn1} APER or \gls{json} and sends it to the dApp over ZMQ. A separate LCM IPC channel dynamically loads and unloads codelets on subscription and unsubscription.}
\vspace{-.3cm}
\end{figure*}
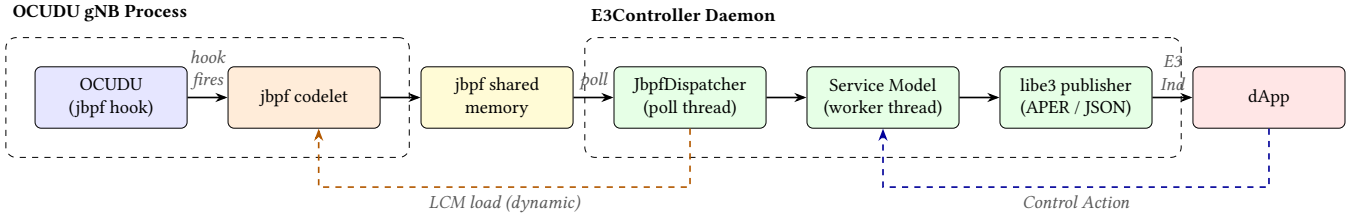


\vspace{-.5cm}

\section{Related Work}

\textbf{dApps and the E3 interface.}
The dApp concept was introduced to extend Open \ran control beyond the latency limits of the \nearrt \ric, first as a vision for co-located \ran applications~\cite{dorodapps2022} and later as a complete architectural framework powered by a new O-RAN logical interface called E3 with a reference implementation on \gls{oai}~\cite{lacavadapps2025}. dApps and E3 are implemented in NVIDIA Aerial~\cite{gpu_isac} and in the Radisys stack~\cite{tiwari2026isac}. A complementary line of work replaces the dApp execution substrate with WebAssembly to provide portable, sandboxed \ran functions~\cite{esper2026enabling}. 

Moreover, several systems address the sub-10\,ms control gap by introducing dedicated real-time controllers rather than co-located applications. EdgeRIC deploys a real-time \ric external to the \ran and supports reinforcement-learning-based scheduling through a custom low-latency interface~\cite{edgeric}, while TinyRIC embeds a lightweight controller and its \emph{tApps} directly within the base station using an in-memory messaging framework~\cite{tinyric}. 
At a different layer of the ecosystem, OAIC provides an open-source platform for developing and evaluating AI-driven \ran-management algorithms~\cite{oaic}. These efforts highlight the need for sub-millisecond application to \ran interaction, while relying on architectures distinct from the emerging O-RAN E3 model~\cite{siva2026ran,doro2026isac,tiwari2026isac}.

\textbf{Fronthaul interception and sensing use cases.}
RANBooster obtains the same I/Q-level data by interposing a kernel-bypass \emph{middlebox} (DPDK/XDP) on the O-RAN fronthaul, which is external and transparent to the \ran~\cite{ranbooster}. We instead capture the same uplink I/Q \emph{in-process} at the O\gls{fh} receive and upper-PHY hooks, avoiding a separate packet-processing path while still exporting the data over a standardized interface. Moreover, hooks at multiple processing stages enable dApps to access and process different classes of I/Q data depending on the application, facilitating the isolation of specific signals and allowing instrumentation to be placed at the most appropriate point in the \ran pipeline. 

\textbf{eBPF-based \ran programmability.}
Janus~\cite {jbpf} pioneered the use of eBPF-style codelets to instrument and control a running \ran by dynamically injecting verified bytecode into a 5G base station to enable low-latency analytics and control. Janus exposes programmability through a bespoke control plane tightly coupled with the underlying \ran implementation. In this work, we build on the \texttt{jbpf} userspace framework and the jbpf-enabled srsRAN fork~\cite{foukas2025srsranjbpf}, extending a newer version of OCUDU with jbpf support while deploying the \texttt{E3Controller} as a sidecar process. This architecture enables the E3 interface without embedding its protocol logic into the \ran implementation.

\textbf{Positioning of this work.}
Existing efforts have independently explored E3-enabled dApps, eBPF-based \ran programmability, real-time control frameworks, and fronthaul-level data extraction. However, to the best of our knowledge, no prior work combines these elements into an end-to-end E3 implementation for OCUDU. This paper fills that gap by extending OCUDU with the \texttt{E3Controller}, a component independent of the protocol stack that enables interoperability with the same dApps that work for NVIDIA Aerial and \gls{oai}. The design bridges jbpf-based instrumentation to standardized E3 procedures through a minimal-touch sidecar architecture, requiring only lightweight hook points within the \ran while leaving the core protocol stack unchanged. The resulting platform exposes fronthaul- and PHY-level data streams to E3-compliant dApps and is validated on a commercial O-RAN 7.2x deployment.

\section{An E3 Controller for Real-Time Open RAN Intelligence}
This section describes the design choices and the architecture behind our \texttt{E3Controller}.\footnote{\url{https://github.com/wineslab/E3Controller}} It provides an overview of how the E3 interface has been designed in the OCUDU context, the various service models supported by this version, how data is distributed among several dApps, and finally, how this introduction may impact the overall performance of the OCUDU \ran software.

\subsection{Sidecar Architecture}
To reduce the impact on the \ran software implementation and make it reusable across other \ran software stacks~\cite{nikaein2014,jbpf}, we decided to separate E3 interface management from the \ran implementation itself. We designed a controller that manages the encoding/decoding of E3-related information (\gls{e3ap} and \gls{e3sm}), dApp subscriptions and controls, and message dispatching as a separate process that runs alongside the \ran software stacks. The integration follows a \emph{sidecar} architecture where the \texttt{E3Controller} runs as a separate process alongside the OCUDU \gls{gnb}. It communicates with the \ran via jbpf's shared-memory \gls{ipc} for data, and via jbpf's \gls{lcm} \gls{ipc} socket for codelet lifecycle management. Consequently, only the hook macros and the jbpf agent initialization need to live inside OCUDU, while the rest of the pipeline (E3 protocol stack, \gls{asn1}/\gls{json} codecs, service models, dApps) sits behind a clean \gls{ipc} boundary and can evolve independently. 

Figure~\ref{fig:flow} summarizes the end-to-end flow. When a dApp subscribes to a service model, the \texttt{E3Controller} \emph{dynamically} allocates the corresponding jbpf codelet into the running \gls{gnb} through the \gls{lcm} \gls{ipc} interface. When the hook is called, the codelet extracts the E3-relevant information and sends it to the jbpf output channel identified by a unique \texttt{stream\_id}. The data is exported through POSIX shared memory to the \texttt{E3Controller}, which is the \gls{ipc} primary. Inside the controller, a service model encodes the data as an \texttt{Indication} message, in \gls{asn1} APER, or \gls{json}, selectable per subscription, and the E3 agent delivers it to subscribers over ZMQ. The matching unload request is issued when the last subscriber leaves.

\subsection{Dispatcher and Three-Stage Pipeline}
\label{sec:pipeline}

The \texttt{E3Controller} is organized as a three-stage pipeline comprising a \emph{poll}, \emph{worker}, and \emph{publisher} stage, each executed on a dedicated thread to minimize latency along the \ran-to-dApp data path.

\noindent \textbf{Poll.} A single \emph{JbpfDispatcher} thread routes each output jbpf buffer to a data-plane pipeline associated with one or more \glspl{sm}, using the buffer's \texttt{stream\_id} as the lookup key. To support high-throughput operation, the dispatcher is deliberately designed as a lightweight component that performs only stream demultiplexing, leaving all payload processing to the pipelines. 

\noindent\textbf{Worker.} Each codelet stream is served by a dedicated pipeline, which acts as the shared data plane for one or more service models. The pipeline drains its \gls{spsc} queue, applies any service-model-specific processing, and forwards the resulting data by reference to the subscribed service models. Each service model then constructs the corresponding \texttt{Indication} and passes it to \texttt{libE3} for transmission (Figure~\ref{fig:flow}). The amount of work performed at this stage is service-model-specific. For instance, the Spectrum \gls{sm} decompresses and remaps each fronthaul section, whereas in the L1 \gls{sm} the worker validates the shared-memory data descriptors and fills in the \texttt{Indication}. In the latter case, the pipeline is a pure metadata path, and the \texttt{E3Controller} never touches I/Q samples.

To maximize throughput, the dispatcher normally releases each jbpf buffer immediately after dispatching it. However, for large payloads from codelets, it may transfer \emph{ownership} of the buffer to the worker thread, which releases it only after all subscribed service models have consumed the data. This allows to eliminate an extra copy of payload while preserving synchronous delivery to all consumers. 

\noindent\textbf{Publisher.} Finally, the \texttt{libE3} publisher thread performs the outer \gls{e3ap} serialization (\gls{asn1} APER or \gls{json}) and transmits the resulting message over ZMQ.

\vspace{-.3cm}
\subsection{Service Models: Spectrum and L1}
\label{sec:sms}

Each \gls{sm} leverages the \texttt{libE3::ServiceModel} interface provided by \texttt{libE3} to leverage the library's \gls{sm} lifecycle management and to expose \gls{sm}-specific metadata, i.e., the \gls{ran} function definition and its telemetry and control identifiers, to the dApps over the \gls{e3ap}. Upon the first dApp subscription, the service model issues an \gls{lcm} load request to attach its codelet to the appropriate \gls{gnb} hook. Conversely, when the last dApp unsubscribes, the service model unloads the codelet, ensuring that instrumentation remains active only when needed. This version of the \texttt{E3Controller} supports two reference service models, \emph{Spectrum}~\cite{lacavadapps2025} and \emph{L1}~\cite{gpu_isac}, which instrument the same uplink I/Q stream at different stages of the \gls{gnb} pipeline while adopting different data-delivery strategies. \emph{Spectrum} hooks the pipeline before the upper-PHY's decompression and exports the captured samples through \gls{asn1}-encoded E3 messages. In contrast, \emph{L1} operates after decompression, where the payloads are substantially larger, and therefore delivers the sample buffers via shared memory to avoid the overhead of encoding them in \gls{asn1}. Together, they represent two complementary points in the hook-placement and service-model design space.

\noindent\textbf{Spectrum SM (per-section, pre-decompression).} The Spectrum \gls{sm} loads a codelet that attaches to the \gls{fh} U-Plane \gls{ul} hook. For each received fronthaul section, the codelet extracts the eCPRI/U-Plane metadata (frame, subframe, slot, symbol, and \gls{prb} range) and writes the compressed I/Q payload to jbpf shared memory. On the \texttt{E3Controller}, the pipeline worker then decompresses each section into int16 I/Q samples, maps subcarriers to FFT bins following the 3GPP convention (guard-band zero-padded), and encodes the Indication messages.

\noindent\textbf{L1 SM (per-slot, post-decompression).} The L1 \gls{sm} loads a codelet that attaches to the upper-PHY hook after the \gls{gnb} has assembled the per-slot resource grid for demodulation. The hook triggers once per slot, on the last symbol, and provides the codelet with a read-only pointer to the complete resource grid. To avoid embedding the large I/Q payload in every E3 \texttt{Indication}, the L1 \gls{sm} stores each captured slot in a shared-memory region, \texttt{/e3\_ran\_buffers}~\cite{gpu_isac} allocated by the \texttt{E3Controller}, and sends an \texttt{Indication} containing only a reference to the corresponding slot. This allows multiple dApps to access the same I/Q data without duplicating or re-encoding the payload.

To transfer the samples into shared memory, the codelet invokes a \emph{jbpf helper function}, registered by the \gls{gnb} during agent initialization. The helper converts the resource grid from \code{bf16} to the \code{fp16} row format required by the dApp and writes it to \texttt{/e3\_ran\_buffers} in a single copy, leaving only a ${\sim}64$\,B descriptor on the jbpf output channel. This delegation reflects the eBPF execution model: the conversion is not expressible in bytecode, while the copy is substantially more efficient when performed by a native routine. The codelet supplies only the source window and a selector, while the helper derives the target row from the buffer geometry maintained by the \texttt{E3Controller}. Thus, the payload bypasses the jbpf ring, which carries only the descriptor containing the slot identity, row indices, and timestamps used in Section~\ref{sec:eval}. 
\subsection{Subscriptions and Data Distribution}
\label{sec:fanout}

\begin{figure}[t]
\centering
\resizebox{\linewidth}{!}{
%
%
\begin{tikzpicture}[
    >=Stealth,
    actor/.style={
        draw, rectangle, rounded corners=2pt,
        minimum width=3.6cm, minimum height=1.2cm,
        align=center, font=\LARGE\bfseries,
        fill=blue!8
    },
    actorgnb/.style={actor, fill=orange!12},
    actordapp/.style={actor, fill=red!10},
    actorctrl/.style={actor, fill=green!10},
    msg/.style={->, thick, line width=0.7pt},
    msglbl/.style={font=\Large, midway, above, fill=white, inner sep=3pt, align=center},
    notelabel/.style={
        font=\Large\itshape,
        text=black!75,
        align=center,
        fill=yellow!15,
        draw=yellow!50!black, line width=0.4pt,
        rounded corners=2pt,
        inner sep=4pt
    },
    lifeline/.style={very thick, dashed, draw=gray!55},
    phaselabel/.style={
        font=\Large\bfseries,
        text=gray!55!black,
        anchor=west
    },
    phasebox/.style={dashed, rounded corners=3pt, draw=gray!55, line width=0.6pt}
]
    \def\xD{0}      
    \def\xC{7.0}    
    \def\xG{14.0}   

    \node[actordapp] (D) at (\xD, 0)  {dApp};
    \node[actorctrl] (C) at (\xC, 0)  {E3Controller\\(sidecar)};
    \node[actorgnb]  (G) at (\xG, 0)  {OCUDU gNB\\(jbpf agent + codelet)};

    \draw[lifeline] (\xD, -0.65) -- (\xD, -11.8);
    \draw[lifeline] (\xC, -0.65) -- (\xC, -11.8);
    \draw[lifeline] (\xG, -0.65) -- (\xG, -11.8);

    \node[phaselabel] at (\xD - 1.4, -1.1) {1. Subscription};
    \draw[msg] (\xD, -2) -- (\xC, -2) node[msglbl] {E3AP \texttt{Subscribe(ranFunctionId)}};
    \draw[msg] (\xC, -2.5) -- (\xG, -2.5) node[msglbl] {LCM \texttt{load(codelet)}};
    \draw[msg] (\xG, -3.3) -- (\xC, -3.3) node[msglbl] {LCM \texttt{ack} (codelet attached to hook)};
    \draw[msg] (\xC, -4.1) -- (\xD, -4.1) node[msglbl] {\texttt{SubscribeResponse}};

    \node[phaselabel] at (\xD - 1.4, -5.1) {2. Data path (\textit{loop, per fire})};
    \draw[phasebox] (\xD - 1.4, -5.3) rectangle (\xG + 2.0, -9.2);

    \node[notelabel] at (\xG, -5.9) {hook fires;\\codelet $\rightarrow$ jbpf SHM};

    \draw[msg] (\xG, -6.9) -- (\xC, -6.9) node[msglbl] {dispatcher drain (jbpf SHM)};

    \node[notelabel] at (\xC, -7.7) {SM worker: process / encode\\libe3 publisher: emit};

    \draw[msg] (\xC, -8.7) -- (\xD, -8.7) node[msglbl] {E3 \texttt{Indication}};

    \node[phaselabel] at (\xD - 1.4, -9.8) {3. Unsubscription};
    \draw[msg] (\xD, -10.6) -- (\xC, -10.6) node[msglbl] {E3AP \texttt{Unsubscribe}};
    \draw[msg] (\xC, -11.0) -- (\xG, -11.0) node[msglbl] {LCM \texttt{unload} (last subscriber)};
    \draw[msg] (\xC, -11.7) -- (\xD, -11.7) node[msglbl] {\texttt{UnsubscribeResponse}};
\end{tikzpicture}}
\caption{Life-cycle of a Single dApp Subscription.}
\label{fig:subscribe-flow}
\vspace{-.7cm}
\end{figure}
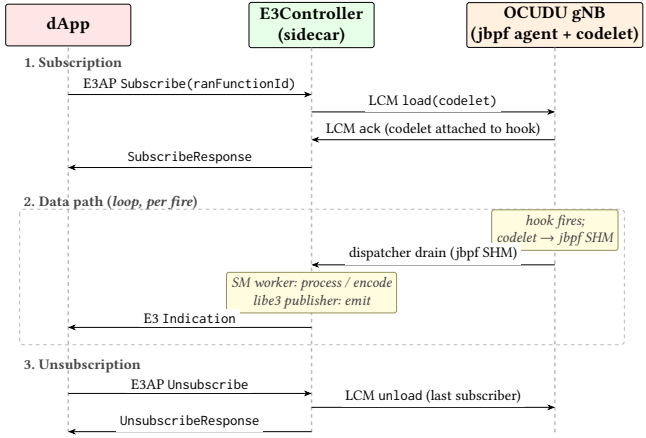

Figure~\ref{fig:subscribe-flow} summarizes the life cycle of a single dApp subscription, consisting of three phases. First, the \texttt{E3Controller} loads the codelet associated with the requested \ran function through the jbpf \gls{lcm} \gls{ipc} channel and acknowledges the subscription. During operation, each codelet writes data to jbpf shared memory, which the \texttt{E3Controller} dispatcher drains and delivers as an E3 \texttt{Indication} to the dApp via \texttt{libE3}. Finally, when the last subscriber unsubscribes, the \texttt{E3Controller} unloads the corresponding codelet. Within this life cycle, the \texttt{E3Controller} scales independently along two orthogonal dimensions: the number of dApps subscribed to a given \ran function and the number of distinct \ran functions served concurrently.

\noindent\textbf{Many dApps, one data type.} The capture chain executes once per codelet invocation, while the service-model-specific processing performed by the pipeline worker is likewise applied only once to each captured sample, regardless of the number of subscribed dApps. The only cost that scales with the number of subscribers is outbound fan-out, whose behavior depends on the service model's indication strategy, as described in the previous section. Thus, in both strategies, in-\ran instrumentation and per-sample processing remain independent of subscriber count. They differ only in whether the publisher replicates the payload itself or distributes a reference to the shared data.


\noindent\textbf{One dApp, many data types.} Each data type is exposed as an independent E3 \emph{\ran function}, implemented through a dedicated {codelet, \texttt{stream\_id}, pipeline, service model} tuple. Consequently, a dApp requiring multiple data types (e.g., uplink I/Q samples and decoded PDUs) establishes separate subscriptions, one for each \ran function. The \texttt{libE3} runtime demultiplexes indications based on the corresponding \texttt{ranFunctionId}, while each service model publishes independently and asynchronously. The dApp correlates across data types using the common identifier carried by all indications. Furthermore, each pipeline is instantiated only when the first subscriber arrives and deleted when the last subscriber unsubscribes, ensuring that inactive \ran functions incur no steady-state overhead.

This design naturally decouples scalability with respect to the number of exported \ran functions from scalability with respect to the number of subscribers. At the jbpf layer, codelets attached to the same hook execute serially within the invoking \ran thread. Therefore, distinct data types are exported via separate hooks at appropriate stages of the \gls{gnb} processing pipeline. Because different \ran threads invoke these hooks, their associated codelets execute independently and do not interfere with one another. Only multiple codelets attached to the same hook are serialized, making hook placement the primary determinant of execution isolation.

\subsection{Minimal Footprint in OCUDU}
\label{sec:footprint}
Enabling the \texttt{E3Controller} on OCUDU requires three modifications to the \gls{gnb} codebase, all of which are conditionally compiled. First, we introduce two jbpf instrumentation hooks into the data path. The first, located in the O\gls{fh} receive path, is ported from Microsoft's jbpf-enabled srsRAN fork~\cite{foukas2025srsranjbpf}. The second, introduced in this work, is placed in the upper-PHY symbol handler, where it exposes a read-only view of the resource grid and triggers a hook invocation on the last symbol of each slot. When jbpf support is disabled, both hooks are compiled out and introduce no runtime overhead. Second, the jbpf agent is initialized as an \gls{ipc} secondary process during \gls{gnb} startup, while the \gls{lcm} \gls{ipc} server is enabled to allow the \texttt{E3Controller} to dynamically load and unload codelets at runtime. The same initialization procedure registers the two helper functions used by the L1 \gls{sm}, namely region attachment and slot publication. These constitute the only \ran-side components that access I/Q samples, and both reuse the conversion routine of the \texttt{E3Controller} without modification, ensuring that the two writers produce byte-identical rows. Finally, the \gls{gnb} configuration is extended with a dedicated \texttt{jbpf:} YAML section, parsed at startup, that specifies the \gls{ipc} namespace, the \gls{lcm} socket path, and the hugepage allocation size.

We have implemented these modifications in a public OCUDU fork based on OCUDU~26.04\footnote{\url{https://github.com/wineslab/ocudu-e3}}.

\begin{figure}[t]
    \centering
    \includegraphics[width=\linewidth]{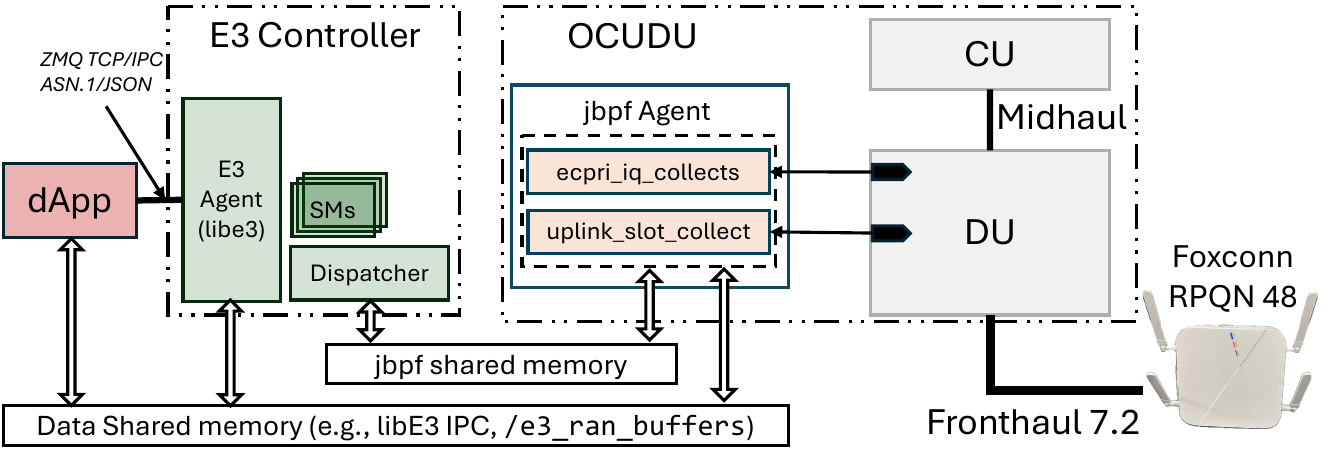}
    \caption{Testbed used for the evaluation.}
    \label{fig:testbed}
    \vspace{-.5cm}
\end{figure}

\begin{figure*}[t]
  \centering
  \newlength{\fwidth}
  \newlength{\fheight}
  \setlength{\fwidth}{\columnwidth}\setlength{\fheight}{0.3\columnwidth}
  \begin{subfigure}{\columnwidth}\centering
\definecolor{darkgray176}{RGB}{176,176,176}
\definecolor{cjson}{RGB}{181,140,198}
\definecolor{casn}{RGB}{127,186,176}
\definecolor{cstd}{RGB}{138,138,138}
\definecolor{coff}{RGB}{201,162,101}
\begin{tikzpicture}
\begin{axis}[
width=\fwidth, height=\fheight,
tick align=outside, tick pos=left, grid style={darkgray176}, ymajorgrids, xmajorgrids,
xlabel style={font=\scriptsize}, ylabel style={font=\scriptsize}, tick label style={font=\scriptsize},
xlabel shift=-3pt, ylabel shift=-3pt, font=\scriptsize,
ylabel={CDF}, ymin=0, ymax=1, ytick={0,0.25,0.5,0.75,1},
legend cell align={left},
reverse legend,
legend style={font=\tiny, fill opacity=0.85, draw=darkgray176, at={(0.5,1.03)}, anchor=south, legend columns=-1},
xlabel={Throughput [Mbps]}, xmin=420, xmax=512,
]
\addplot[cjson, line width=0.9pt, mark=triangle*, mark size=1.3pt, mark repeat=12, mark phase=10, mark options={solid}] coordinates {(275.0,0.0033) (368.0,0.0133) (434.0,0.0233) (446.0,0.0300) (454.0,0.0400) (459.0,0.0500) (464.0,0.0600) (465.0,0.0700) (466.0,0.0800) (466.0,0.0867) (467.0,0.0967) (468.0,0.1067) (468.0,0.1167) (469.0,0.1267) (470.0,0.1333) (471.0,0.1433) (472.0,0.1533) (472.0,0.1633) (473.0,0.1733) (473.0,0.1833) (474.0,0.1900) (475.0,0.2000) (476.0,0.2100) (478.0,0.2200) (478.0,0.2300) (479.0,0.2400) (480.0,0.2467) (480.0,0.2567) (480.0,0.2667) (482.0,0.2767) (482.0,0.2867) (483.0,0.2933) (484.0,0.3033) (485.0,0.3133) (486.0,0.3233) (487.0,0.3333) (487.0,0.3433) (487.0,0.3500) (488.0,0.3600) (489.0,0.3700) (490.0,0.3800) (490.0,0.3900) (490.0,0.3967) (491.0,0.4067) (491.0,0.4167) (491.0,0.4267) (492.0,0.4367) (492.0,0.4467) (493.0,0.4533) (493.0,0.4633) (493.0,0.4733) (493.0,0.4833) (493.0,0.4933) (494.0,0.5033) (494.0,0.5100) (494.0,0.5200) (494.0,0.5300) (494.0,0.5400) (494.0,0.5500) (494.0,0.5567) (495.0,0.5667) (495.0,0.5767) (495.0,0.5867) (495.0,0.5967) (495.0,0.6067) (495.0,0.6133) (495.0,0.6233) (495.0,0.6333) (496.0,0.6433) (496.0,0.6533) (496.0,0.6600) (496.0,0.6700) (496.0,0.6800) (496.0,0.6900) (496.0,0.7000) (496.0,0.7100) (496.0,0.7167) (497.0,0.7267) (497.0,0.7367) (497.0,0.7467) (497.0,0.7567) (497.0,0.7633) (497.0,0.7733) (497.0,0.7833) (497.0,0.7933) (498.0,0.8033) (498.0,0.8133) (498.0,0.8200) (498.0,0.8300) (498.0,0.8400) (498.0,0.8500) (498.0,0.8600) (498.0,0.8700) (498.0,0.8767) (498.0,0.8867) (498.0,0.8967) (498.0,0.9067) (499.0,0.9167) (499.0,0.9233) (499.0,0.9333) (499.0,0.9433) (499.0,0.9533) (500.0,0.9633) (500.0,0.9733) (500.0,0.9800) (501.0,0.9900) (508.0,1.0000)};
\addlegendentry{JSON}
\addplot[casn, line width=0.9pt, mark=*, mark size=1.1pt, mark repeat=12, mark phase=6, mark options={solid}] coordinates {(426.0,0.0033) (470.0,0.0133) (473.0,0.0233) (473.0,0.0300) (475.0,0.0400) (476.0,0.0500) (478.0,0.0600) (478.0,0.0700) (479.0,0.0800) (480.0,0.0867) (480.0,0.0967) (481.0,0.1067) (482.0,0.1167) (483.0,0.1267) (483.0,0.1333) (484.0,0.1433) (485.0,0.1533) (487.0,0.1633) (487.0,0.1733) (488.0,0.1833) (488.0,0.1900) (488.0,0.2000) (489.0,0.2100) (489.0,0.2200) (489.0,0.2300) (489.0,0.2400) (490.0,0.2467) (490.0,0.2567) (490.0,0.2667) (491.0,0.2767) (491.0,0.2867) (491.0,0.2933) (492.0,0.3033) (492.0,0.3133) (492.0,0.3233) (492.0,0.3333) (492.0,0.3433) (492.0,0.3500) (492.0,0.3600) (493.0,0.3700) (493.0,0.3800) (493.0,0.3900) (493.0,0.3967) (493.0,0.4067) (493.0,0.4167) (493.0,0.4267) (494.0,0.4367) (494.0,0.4467) (494.0,0.4533) (494.0,0.4633) (494.0,0.4733) (494.0,0.4833) (494.0,0.4933) (495.0,0.5033) (495.0,0.5100) (495.0,0.5200) (495.0,0.5300) (495.0,0.5400) (495.0,0.5500) (495.0,0.5567) (495.0,0.5667) (495.0,0.5767) (495.0,0.5867) (496.0,0.5967) (496.0,0.6067) (496.0,0.6133) (496.0,0.6233) (496.0,0.6333) (496.0,0.6433) (496.0,0.6533) (496.0,0.6600) (496.0,0.6700) (496.0,0.6800) (496.0,0.6900) (496.0,0.7000) (496.0,0.7100) (496.0,0.7167) (497.0,0.7267) (497.0,0.7367) (497.0,0.7467) (497.0,0.7567) (497.0,0.7633) (497.0,0.7733) (497.0,0.7833) (497.0,0.7933) (497.0,0.8033) (497.0,0.8133) (497.0,0.8200) (498.0,0.8300) (498.0,0.8400) (498.0,0.8500) (498.0,0.8600) (498.0,0.8700) (498.0,0.8767) (498.0,0.8867) (498.0,0.8967) (498.0,0.9067) (499.0,0.9167) (499.0,0.9233) (499.0,0.9333) (499.0,0.9433) (499.0,0.9533) (500.0,0.9633) (501.0,0.9733) (501.0,0.9800) (502.0,0.9900) (512.0,1.0000)};
\addlegendentry{Asn1}
\addplot[cstd, line width=0.9pt, mark=square*, mark size=1.0pt, mark repeat=12, mark phase=2, mark options={solid}] coordinates {(437.0,0.0033) (452.0,0.0133) (459.0,0.0233) (466.0,0.0300) (468.0,0.0400) (469.0,0.0500) (470.0,0.0600) (470.0,0.0700) (471.0,0.0800) (473.0,0.0867) (473.0,0.0967) (474.0,0.1067) (475.0,0.1167) (476.0,0.1267) (477.0,0.1333) (478.0,0.1433) (478.0,0.1533) (478.0,0.1633) (479.0,0.1733) (479.0,0.1833) (479.0,0.1900) (480.0,0.2000) (480.0,0.2100) (481.0,0.2200) (481.0,0.2300) (481.0,0.2400) (481.0,0.2467) (482.0,0.2567) (482.0,0.2667) (482.0,0.2767) (483.0,0.2867) (483.0,0.2933) (483.0,0.3033) (484.0,0.3133) (484.0,0.3233) (484.0,0.3333) (485.0,0.3433) (485.0,0.3500) (485.0,0.3600) (486.0,0.3700) (486.0,0.3800) (487.0,0.3900) (487.0,0.3967) (487.0,0.4067) (488.0,0.4167) (488.0,0.4267) (488.0,0.4367) (489.0,0.4467) (489.0,0.4533) (489.0,0.4633) (489.0,0.4733) (490.0,0.4833) (490.0,0.4933) (491.0,0.5033) (491.0,0.5100) (491.0,0.5200) (491.0,0.5300) (492.0,0.5400) (492.0,0.5500) (492.0,0.5567) (492.0,0.5667) (493.0,0.5767) (493.0,0.5867) (493.0,0.5967) (494.0,0.6067) (494.0,0.6133) (494.0,0.6233) (494.0,0.6333) (494.0,0.6433) (495.0,0.6533) (495.0,0.6600) (495.0,0.6700) (495.0,0.6800) (495.0,0.6900) (495.0,0.7000) (495.0,0.7100) (496.0,0.7167) (496.0,0.7267) (496.0,0.7367) (496.0,0.7467) (496.0,0.7567) (497.0,0.7633) (497.0,0.7733) (497.0,0.7833) (497.0,0.7933) (497.0,0.8033) (497.0,0.8133) (497.0,0.8200) (498.0,0.8300) (498.0,0.8400) (498.0,0.8500) (498.0,0.8600) (498.0,0.8700) (498.0,0.8767) (498.0,0.8867) (498.0,0.8967) (498.0,0.9067) (499.0,0.9167) (499.0,0.9233) (499.0,0.9333) (499.0,0.9433) (499.0,0.9533) (500.0,0.9633) (500.0,0.9733) (501.0,0.9800) (508.0,0.9900) (531.0,1.0000)};
\addlegendentry{Standard}
\addplot[coff, line width=0.9pt] coordinates {(441.0,0.0033) (475.0,0.0133) (478.0,0.0233) (479.0,0.0300) (479.0,0.0400) (479.0,0.0500) (480.0,0.0600) (480.0,0.0700) (481.0,0.0800) (482.0,0.0867) (484.0,0.0967) (484.0,0.1067) (485.0,0.1167) (485.0,0.1267) (486.0,0.1333) (487.0,0.1433) (487.0,0.1533) (488.0,0.1633) (488.0,0.1733) (489.0,0.1833) (489.0,0.1900) (489.0,0.2000) (490.0,0.2100) (490.0,0.2200) (490.0,0.2300) (491.0,0.2400) (491.0,0.2467) (491.0,0.2567) (492.0,0.2667) (492.0,0.2767) (492.0,0.2867) (492.0,0.2933) (493.0,0.3033) (493.0,0.3133) (493.0,0.3233) (493.0,0.3333) (493.0,0.3433) (494.0,0.3500) (494.0,0.3600) (494.0,0.3700) (494.0,0.3800) (494.0,0.3900) (494.0,0.3967) (494.0,0.4067) (494.0,0.4167) (495.0,0.4267) (495.0,0.4367) (495.0,0.4467) (495.0,0.4533) (495.0,0.4633) (495.0,0.4733) (495.0,0.4833) (495.0,0.4933) (496.0,0.5033) (496.0,0.5100) (496.0,0.5200) (496.0,0.5300) (496.0,0.5400) (496.0,0.5500) (496.0,0.5567) (496.0,0.5667) (496.0,0.5767) (497.0,0.5867) (497.0,0.5967) (497.0,0.6067) (497.0,0.6133) (497.0,0.6233) (497.0,0.6333) (497.0,0.6433) (497.0,0.6533) (497.0,0.6600) (497.0,0.6700) (497.0,0.6800) (497.0,0.6900) (497.0,0.7000) (497.0,0.7100) (497.0,0.7167) (497.0,0.7267) (498.0,0.7367) (498.0,0.7467) (498.0,0.7567) (498.0,0.7633) (498.0,0.7733) (498.0,0.7833) (498.0,0.7933) (498.0,0.8033) (498.0,0.8133) (498.0,0.8200) (499.0,0.8300) (499.0,0.8400) (499.0,0.8500) (499.0,0.8600) (499.0,0.8700) (499.0,0.8767) (499.0,0.8867) (499.0,0.8967) (499.0,0.9067) (500.0,0.9167) (500.0,0.9233) (500.0,0.9333) (500.0,0.9433) (500.0,0.9533) (501.0,0.9633) (502.0,0.9733) (502.0,0.9800) (506.0,0.9900) (517.0,1.0000)};
\addlegendentry{Official}
\end{axis}
\end{tikzpicture}
    \caption{\gls{dl} throughput}
    \label{fig:cdf_dl}
    \Description{Empirical cumulative distribution function of the per-second downlink \gls{udp} throughput pooled across runs. The three curves for the Standard baseline, the \gls{asn1}-encoded dApp, and the \gls{json}-encoded dApp are nearly indistinguishable, all reaching approximately 490 Mbps, indicating no measurable downlink throughput impact from the \texttt{E3Controller}.}
  \end{subfigure}%
  \begin{subfigure}{\columnwidth}\centering
\definecolor{darkgray176}{RGB}{176,176,176}
\definecolor{cjson}{RGB}{181,140,198}
\definecolor{casn}{RGB}{127,186,176}
\definecolor{cstd}{RGB}{138,138,138}
\definecolor{coff}{RGB}{201,162,101}
\begin{tikzpicture}
\begin{axis}[
width=\fwidth, height=\fheight,
tick align=outside, tick pos=left, grid style={darkgray176}, ymajorgrids, xmajorgrids,
xlabel style={font=\scriptsize}, ylabel style={font=\scriptsize}, tick label style={font=\scriptsize},
xlabel shift=-3pt, ylabel shift=-3pt, font=\scriptsize,
ylabel={CDF}, ymin=0, ymax=1, ytick={0,0.25,0.5,0.75,1},
legend cell align={left},
reverse legend,
legend style={font=\tiny, fill opacity=0.85, draw=darkgray176, at={(0.5,1.03)}, anchor=south, legend columns=-1},
xlabel={Throughput [Mbps]}, xmin=60, xmax=80,
]
\addplot[cjson, line width=0.9pt, mark=triangle*, mark size=1.3pt, mark repeat=12, mark phase=10, mark options={solid}] coordinates {(48.8,0.0033) (58.0,0.0133) (59.7,0.0233) (60.7,0.0300) (61.8,0.0400) (63.7,0.0500) (64.7,0.0600) (65.7,0.0700) (67.4,0.0800) (67.9,0.0867) (68.1,0.0967) (69.0,0.1067) (69.6,0.1167) (70.0,0.1267) (70.5,0.1333) (70.7,0.1433) (70.9,0.1533) (71.1,0.1633) (71.4,0.1733) (71.6,0.1833) (71.7,0.1900) (71.8,0.2000) (72.0,0.2100) (72.4,0.2200) (72.5,0.2300) (72.7,0.2400) (72.8,0.2467) (72.8,0.2567) (73.0,0.2667) (73.2,0.2767) (73.3,0.2867) (73.4,0.2933) (73.5,0.3033) (73.5,0.3133) (73.6,0.3233) (73.7,0.3333) (73.7,0.3433) (73.7,0.3500) (73.8,0.3600) (73.8,0.3700) (73.9,0.3800) (73.9,0.3900) (74.0,0.3967) (74.0,0.4067) (74.0,0.4167) (74.1,0.4267) (74.1,0.4367) (74.2,0.4467) (74.2,0.4533) (74.2,0.4633) (74.2,0.4733) (74.4,0.4833) (74.4,0.4933) (74.4,0.5033) (74.4,0.5100) (74.5,0.5200) (74.5,0.5300) (74.5,0.5400) (74.6,0.5500) (74.6,0.5567) (74.6,0.5667) (74.6,0.5767) (74.7,0.5867) (74.7,0.5967) (74.7,0.6067) (74.7,0.6133) (74.7,0.6233) (74.8,0.6333) (74.8,0.6433) (74.9,0.6533) (74.9,0.6600) (74.9,0.6700) (75.0,0.6800) (75.0,0.6900) (75.0,0.7000) (75.0,0.7100) (75.1,0.7167) (75.1,0.7267) (75.2,0.7367) (75.2,0.7467) (75.2,0.7567) (75.2,0.7633) (75.2,0.7733) (75.2,0.7833) (75.3,0.7933) (75.3,0.8033) (75.4,0.8133) (75.4,0.8200) (75.4,0.8300) (75.4,0.8400) (75.5,0.8500) (75.5,0.8600) (75.6,0.8700) (75.6,0.8767) (75.6,0.8867) (75.7,0.8967) (75.7,0.9067) (75.8,0.9167) (75.8,0.9233) (75.9,0.9333) (76.0,0.9433) (76.0,0.9533) (76.0,0.9633) (76.1,0.9733) (76.2,0.9800) (76.5,0.9900) (77.9,1.0000)};
\addlegendentry{JSON}
\addplot[casn, line width=0.9pt, mark=*, mark size=1.1pt, mark repeat=12, mark phase=6, mark options={solid}] coordinates {(48.3,0.0033) (54.0,0.0133) (57.7,0.0233) (58.9,0.0300) (61.0,0.0400) (62.3,0.0500) (63.1,0.0600) (63.9,0.0700) (64.7,0.0800) (65.2,0.0867) (65.8,0.0967) (66.2,0.1067) (66.5,0.1167) (67.3,0.1267) (67.4,0.1333) (67.5,0.1433) (68.2,0.1533) (68.9,0.1633) (69.2,0.1733) (69.7,0.1833) (70.0,0.1900) (70.2,0.2000) (70.8,0.2100) (71.0,0.2200) (71.2,0.2300) (71.3,0.2400) (71.5,0.2467) (71.6,0.2567) (71.7,0.2667) (71.9,0.2767) (72.1,0.2867) (72.2,0.2933) (72.3,0.3033) (72.5,0.3133) (72.6,0.3233) (72.8,0.3333) (72.8,0.3433) (73.0,0.3500) (73.1,0.3600) (73.1,0.3700) (73.2,0.3800) (73.3,0.3900) (73.3,0.3967) (73.4,0.4067) (73.6,0.4167) (73.8,0.4267) (73.9,0.4367) (73.9,0.4467) (74.0,0.4533) (74.0,0.4633) (74.0,0.4733) (74.1,0.4833) (74.2,0.4933) (74.2,0.5033) (74.3,0.5100) (74.3,0.5200) (74.4,0.5300) (74.4,0.5400) (74.5,0.5500) (74.5,0.5567) (74.6,0.5667) (74.6,0.5767) (74.7,0.5867) (74.7,0.5967) (74.8,0.6067) (74.8,0.6133) (74.9,0.6233) (75.0,0.6333) (75.0,0.6433) (75.0,0.6533) (75.0,0.6600) (75.0,0.6700) (75.0,0.6800) (75.1,0.6900) (75.1,0.7000) (75.1,0.7100) (75.2,0.7167) (75.2,0.7267) (75.2,0.7367) (75.2,0.7467) (75.2,0.7567) (75.3,0.7633) (75.3,0.7733) (75.3,0.7833) (75.4,0.7933) (75.5,0.8033) (75.5,0.8133) (75.5,0.8200) (75.5,0.8300) (75.6,0.8400) (75.6,0.8500) (75.6,0.8600) (75.6,0.8700) (75.7,0.8767) (75.7,0.8867) (75.8,0.8967) (75.8,0.9067) (75.9,0.9167) (75.9,0.9233) (76.0,0.9333) (76.0,0.9433) (76.1,0.9533) (76.1,0.9633) (76.2,0.9733) (76.3,0.9800) (76.7,0.9900) (78.0,1.0000)};
\addlegendentry{Asn1}
\addplot[cstd, line width=0.9pt, mark=square*, mark size=1.0pt, mark repeat=12, mark phase=2, mark options={solid}] coordinates {(51.3,0.0033) (57.3,0.0133) (60.2,0.0233) (60.4,0.0300) (61.3,0.0400) (63.6,0.0500) (64.2,0.0600) (64.4,0.0700) (64.9,0.0800) (65.0,0.0867) (65.2,0.0967) (65.4,0.1067) (65.8,0.1167) (66.2,0.1267) (66.3,0.1333) (66.6,0.1433) (67.0,0.1533) (67.2,0.1633) (67.6,0.1733) (67.7,0.1833) (67.9,0.1900) (68.2,0.2000) (68.4,0.2100) (68.5,0.2200) (68.9,0.2300) (69.0,0.2400) (69.3,0.2467) (69.5,0.2567) (69.8,0.2667) (70.0,0.2767) (70.3,0.2867) (70.6,0.2933) (70.7,0.3033) (71.0,0.3133) (71.0,0.3233) (71.1,0.3333) (71.4,0.3433) (71.5,0.3500) (71.6,0.3600) (71.7,0.3700) (71.9,0.3800) (72.3,0.3900) (72.3,0.3967) (72.3,0.4067) (72.6,0.4167) (72.8,0.4267) (72.9,0.4367) (73.0,0.4467) (73.1,0.4533) (73.2,0.4633) (73.3,0.4733) (73.3,0.4833) (73.4,0.4933) (73.5,0.5033) (73.5,0.5100) (73.6,0.5200) (73.7,0.5300) (73.8,0.5400) (73.9,0.5500) (74.0,0.5567) (74.0,0.5667) (74.1,0.5767) (74.2,0.5867) (74.2,0.5967) (74.3,0.6067) (74.3,0.6133) (74.4,0.6233) (74.4,0.6333) (74.5,0.6433) (74.5,0.6533) (74.5,0.6600) (74.6,0.6700) (74.7,0.6800) (74.7,0.6900) (74.8,0.7000) (74.9,0.7100) (74.9,0.7167) (74.9,0.7267) (74.9,0.7367) (75.0,0.7467) (75.0,0.7567) (75.0,0.7633) (75.1,0.7733) (75.1,0.7833) (75.2,0.7933) (75.3,0.8033) (75.4,0.8133) (75.4,0.8200) (75.4,0.8300) (75.4,0.8400) (75.5,0.8500) (75.5,0.8600) (75.6,0.8700) (75.6,0.8767) (75.6,0.8867) (75.6,0.8967) (75.7,0.9067) (75.8,0.9167) (75.8,0.9233) (75.8,0.9333) (75.9,0.9433) (75.9,0.9533) (76.0,0.9633) (76.1,0.9733) (76.2,0.9800) (76.4,0.9900) (77.0,1.0000)};
\addlegendentry{Standard}
\addplot[coff, line width=0.9pt] coordinates {(44.0,0.0033) (48.0,0.0133) (54.4,0.0233) (56.8,0.0300) (57.4,0.0400) (58.5,0.0500) (59.0,0.0600) (60.6,0.0700) (62.3,0.0800) (63.1,0.0867) (63.8,0.0967) (65.3,0.1067) (65.9,0.1167) (66.2,0.1267) (66.4,0.1333) (67.3,0.1433) (68.3,0.1533) (68.7,0.1633) (69.3,0.1733) (69.6,0.1833) (69.7,0.1900) (70.0,0.2000) (70.7,0.2100) (70.9,0.2200) (71.3,0.2300) (71.4,0.2400) (72.0,0.2467) (72.0,0.2567) (72.1,0.2667) (72.5,0.2767) (72.7,0.2867) (72.7,0.2933) (72.8,0.3033) (73.0,0.3133) (73.1,0.3233) (73.2,0.3333) (73.2,0.3433) (73.3,0.3500) (73.3,0.3600) (73.4,0.3700) (73.5,0.3800) (73.5,0.3900) (73.5,0.3967) (73.6,0.4067) (73.7,0.4167) (73.7,0.4267) (73.8,0.4367) (73.9,0.4467) (73.9,0.4533) (73.9,0.4633) (73.9,0.4733) (74.0,0.4833) (74.0,0.4933) (74.1,0.5033) (74.1,0.5100) (74.1,0.5200) (74.2,0.5300) (74.3,0.5400) (74.3,0.5500) (74.3,0.5567) (74.4,0.5667) (74.4,0.5767) (74.4,0.5867) (74.5,0.5967) (74.5,0.6067) (74.6,0.6133) (74.6,0.6233) (74.7,0.6333) (74.7,0.6433) (74.7,0.6533) (74.7,0.6600) (74.8,0.6700) (74.8,0.6800) (74.8,0.6900) (74.9,0.7000) (74.9,0.7100) (74.9,0.7167) (74.9,0.7267) (75.0,0.7367) (75.0,0.7467) (75.1,0.7567) (75.1,0.7633) (75.2,0.7733) (75.2,0.7833) (75.2,0.7933) (75.3,0.8033) (75.3,0.8133) (75.3,0.8200) (75.3,0.8300) (75.4,0.8400) (75.4,0.8500) (75.5,0.8600) (75.5,0.8700) (75.6,0.8767) (75.6,0.8867) (75.6,0.8967) (75.7,0.9067) (75.7,0.9167) (75.8,0.9233) (75.8,0.9333) (75.9,0.9433) (76.0,0.9533) (76.1,0.9633) (76.3,0.9733) (76.3,0.9800) (76.6,0.9900) (78.2,1.0000)};
\addlegendentry{Official}
\end{axis}
\end{tikzpicture}
    \caption{\gls{ul} throughput}
    \label{fig:cdf_ul}
    \Description{Empirical cumulative distribution function of the per-second uplink \gls{udp} throughput pooled across runs. The three curves for the Standard baseline, the \gls{asn1}-encoded dApp, and the \gls{json}-encoded dApp overlap, all sustaining approximately 72 Mbps, indicating no measurable uplink throughput impact from the \texttt{E3Controller}.}
  \end{subfigure}
  \caption{Empirical CDF of the UDP throughput for the OCUDU baseline with and without the \texttt{E3Controller}.}
  \label{fig:cdf}
  \Description{Two side-by-side plots of empirical cumulative distribution functions of \gls{udp} throughput, downlink on the left and uplink on the right. In both directions, the curves for the Standard baseline, the \gls{asn1} encoding, and the \gls{json} encoding are indistinguishable within run-to-run variability, showing that exporting fronthaul I/Q samples to dApps through the E3 interface introduces no measurable impact on OCUDU throughput regardless of the indication encoding.}
\end{figure*}
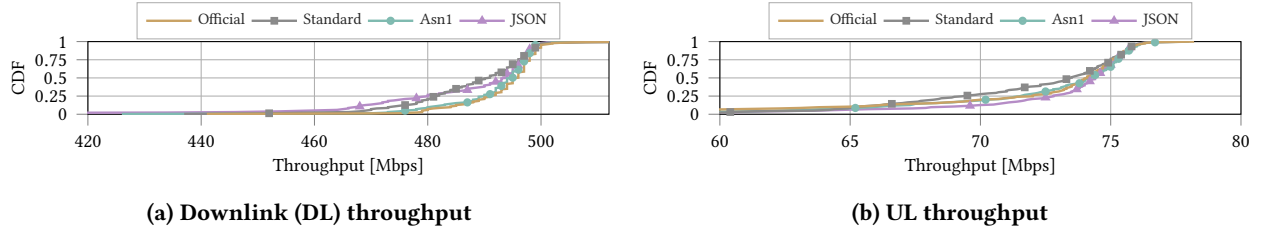

\section{E3Controller Validation and Evaluation}
\label{sec:eval}

In this section, we compare the performance of the OCUDU gNB with and without the \texttt{E3Controller}. Moreover, given the low-latency requirements of the dApp-based scenarios, we analyze the latency introduced by the \texttt{E3Controller} and its components when delivering data to a dApp under high traffic loads.

Figure~\ref{fig:testbed} depicts the O-RAN 7.2x experimental setup, built on the X5G testbed~\cite{villa2025x5g} and the AutoRAN framework~\cite{maxentiautoran2026}, used to evaluate the \texttt{E3Controller}. An OCUDU \gls{gnb}, deployed as a pod on OpenShift, is connected to a Foxconn RPQN~4800 \gls{ru} through a switch. The cell operates in band n78 with $30$~kHz subcarrier spacing and $100$\,MHz channel bandwidth, corresponding to $273$\,\glspl{prb}. The \gls{tdd} configuration follows a 7DS2U pattern, while the fronthaul transports BFP-9 compressed I/Q samples on both the \gls{ul} and \gls{dl}, using 4 \gls{dl} and 4 \gls{ul} antenna ports. We use one Samsung S23 phone as the \gls{ue}, and keep the phone's location constant throughout all experiments. The \texttt{E3Controller} executes within the same pod as the OCUDU process, with its polling, worker, and publisher threads pinned to dedicated CPU cores to minimize scheduling interference. 

The evaluation focuses on the L1 \gls{sm}, as it exercises the \texttt{E3Controller} at its highest sustained data rate. We compare four configurations: the OCUDU~26.04 baseline (\emph{Official}); our fork, synchronized with the corresponding official OCUDU commit and including the two jbpf hooks described in Section~\ref{sec:sms}, but with the hooks excluded from the build (\emph{Standard}); and our full implementation, including both the jbpf hooks and the \texttt{E3Controller}, with the latter streaming I/Q samples to a spectrum-sensing dApp using either \gls{asn1} or \gls{json} for indication encoding. For each, we measure (a) end-to-end throughput with and without the \texttt{E3Controller} and dApp attached alongside the overhead compute and energy cost of the framework, and (b) per-stage latency along the \ran-to-dApp path.

\vspace{-.5cm}

\subsection{Throughput and Energy Cost}
We drive \gls{udp} traffic with \texttt{iperf3} for 60\,s at a target rate of 500\,Mbps in the \gls{dl} and 100\,Mbps in the \gls{ul}, repeating each run five times. Figure~\ref{fig:cdf} reports the empirical \gls{cdf} of the per-second throughput pooled across the five runs. 

In the \gls{dl} direction, all four configurations achieve approximately 490\,Mbps, with nearly identical distributions. Similarly, in the \gls{ul} direction, all configurations sustain approximately 72\,Mbps. The throughput distributions of both the \gls{asn1} and \gls{json} configurations are indistinguishable from the \emph{Official} and \emph{Standard} baselines, indicating that enabling the controller hooks, exporting fronthaul I/Q samples to dApps through the E3 interface, and encoding choice introduces no measurable impact on OCUDU \ran throughput.

The reference dApp served by our \texttt{E3Controller} performs spectrum sensing, introducing a concurrent computational workload whose power consumption is relevant to the evaluation. While the \texttt{E3Controller} and dApp leave throughput unaffected, they consume CPU and energy by processing fronthaul I/Q. We measure the container's power consumption using Kepler, following an energy-profiling methodology for Open RAN deployments~\cite{tenoran}. We report \emph{dynamic} power (the traffic-attributable term isolated from the static platform draw) and CPU occupancy in cores, averaged over the duration of \gls{ota} experiments. 
Table~\ref{tab:energy} summarizes the results. The \emph{Dynamic power} column is the average Kepler dynamic component. The \emph{CPU} column is the container's aggregate CPU occupancy in cores, where each core represents a fully utilized CPU. The last row is the pod floor (\gls{gnb} off); the remaining rows report the four configurations under the 500/100\,Mbps \gls{dl}/\gls{ul} \texttt{iperf3} load.

Compared with the OCUDU baselines, the \texttt{E3Controller} adds roughly two CPU cores through its polling, worker, and publisher threads, with the dApp contributing a smaller, traffic-gated share on top. These threads are statically pinned and poll continuously to guarantee deterministic real-time performance, avoiding context switches and scheduling latency along the critical data path. Because these threads poll rather than compute, the dominant cost of the framework is therefore CPU occupancy, not energy. The stack draws only $14$\,W in the \gls{dl} and $11$\,W in the \gls{ul}, against $11$ and $10$\,W for the baselines, an overhead of a few watts, and its resident memory grows from $3.5$ to $4.4$\,GB, about $0.4$\,GB each for the \texttt{E3Controller} and the dApp. The cost is also direction-independent. The stack occupies $4$\,cores in both the \gls{dl} and \gls{ul}, and the small power difference between them follows the \ran's own load rather than the sensing pipeline. This matches the expected behavior of an \gls{isac} workload, whose footprint tracks the sensed bandwidth rather than the served traffic. The two indication encodings are near-equivalent. They deliver the same throughput (Figure~\ref{fig:cdf}) and draw comparable CPU and dynamic power, differing by about $1$\,W in either direction. 

\begin{table}[t]
\centering
\vspace{-10pt}
\caption{Average dynamic power and CPU occupancy across operating states. Values are reported as mean $\pm$ standard deviation.}
\label{tab:energy}
\footnotesize
\resizebox{\columnwidth}{!}{%
\begin{tabular}{l l c c}
\toprule
State & Configuration & {Dynamic power [\si{\watt}]} & {CPU [cores]} \\
\midrule
\multirow{4}{*}{\gls{dl} 500\,Mbps}
  & \emph{Official}       & $12.4 \pm 1.1$  & $2.12 \pm 0.18$ \\
  & \emph{Standard}       & $11.0 \pm 1.0$  & $2.05 \pm 0.30$ \\
  & E3+dApp (\gls{asn1})  & $13.9 \pm 1.2$  & $4.09 \pm 0.15$ \\
  & E3+dApp (\gls{json})  & $14.3 \pm 0.7$  & $4.28 \pm 0.40$ \\
\midrule
\multirow{4}{*}{\gls{ul} 100\,Mbps}
  & \emph{Official}       & $10.5 \pm 0.1$  & $1.97 \pm 0.09$ \\
  & \emph{Standard}       & $9.7 \pm 0.8$   & $1.85 \pm 0.19$ \\
  & E3+dApp (\gls{asn1})  & $11.3 \pm 1.4$  & $3.72 \pm 0.57$ \\
  & E3+dApp (\gls{json})  & $10.5 \pm 0.3$  & $3.72 \pm 0.22$ \\
\midrule
Pod floor (\gls{gnb} off) & --- & $0.96 \pm 0.01$ & $0.05 \pm 0.01$ \\
\bottomrule
\end{tabular}%
}
\end{table}

\subsection{E3Controller Latency}

A real-time E3 data path must deliver \ran data to the dApp within the sub-millisecond control-loop budget that motivates co-locating dApps with the \ran. To evaluate this requirement, we instrument the complete L1 \gls{sm} pipeline, timestamping every indication from the instant the jbpf hook fires inside OCUDU until its transmission through \texttt{libE3}. We split the end-to-end latency into two components: (i) the \ran-to-controller transfer, including the L1 \gls{sm} processing, and (ii) the publisher-side transmission through \texttt{libE3}.

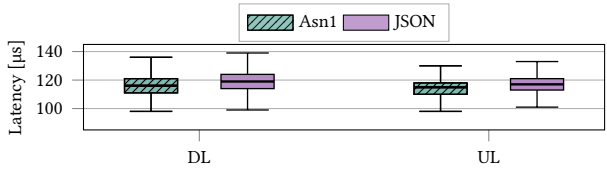
\begin{figure}[t]
  \centering
  \setlength{\fwidth}{\columnwidth}\setlength{\fheight}{0.32\columnwidth}
  \definecolor{darkgray176}{RGB}{176,176,176}
\definecolor{casn}{RGB}{127,186,176}
\definecolor{cjson}{RGB}{181,140,198}
\begin{tikzpicture}
\begin{axis}[
width=\fwidth,
height=\fheight,
font=\scriptsize,
legend cell align={left},
legend style={font=\scriptsize, fill opacity=0.85, draw opacity=1, text opacity=1, draw=darkgray176, at={(0.5,1.03)}, anchor=south, legend columns=-1},
tick align=outside,
tick pos=left,
grid style={darkgray176},
ymajorgrids,
ylabel style={font=\scriptsize},
tick label style={font=\scriptsize},
ylabel shift=-3pt,
ylabel={Latency [\si{\micro\second}]},
boxplot/draw direction=y,
boxplot/box extend=0.55,
boxplot/every median/.style={black, line width=0.9pt},
xtick={1.5,4.5},
xticklabels={\gls{dl},\gls{ul}},
xmin=0.3, xmax=5.7,
ymin=85, ymax=145,
]
\addlegendimage{area legend, fill=casn, draw=black, postaction={pattern=north east lines, pattern color=black}}
\addlegendentry{Asn1}
\addlegendimage{area legend, fill=cjson, draw=black}
\addlegendentry{JSON}
\addplot[boxplot prepared={draw position=1, lower whisker=98, lower quartile=111, median=116, upper quartile=121, upper whisker=136},
  fill=casn, draw=black, line width=0.5pt, forget plot] coordinates {};
\addplot[boxplot prepared={draw position=1, lower whisker=98, lower quartile=111, median=116, upper quartile=121, upper whisker=136},
  pattern=north east lines, pattern color=black, draw=black, line width=0.5pt, forget plot] coordinates {};
\addplot[boxplot prepared={draw position=2, lower whisker=99, lower quartile=114, median=119, upper quartile=124, upper whisker=139},
  fill=cjson, draw=black, line width=0.5pt, forget plot] coordinates {};
\addplot[boxplot prepared={draw position=4, lower whisker=98, lower quartile=110, median=115, upper quartile=118, upper whisker=130},
  fill=casn, draw=black, line width=0.5pt, forget plot] coordinates {};
\addplot[boxplot prepared={draw position=4, lower whisker=98, lower quartile=110, median=115, upper quartile=118, upper whisker=130},
  pattern=north east lines, pattern color=black, draw=black, line width=0.5pt, forget plot] coordinates {};
\addplot[boxplot prepared={draw position=5, lower whisker=101, lower quartile=113, median=117, upper quartile=121, upper whisker=133},
  fill=cjson, draw=black, line width=0.5pt, forget plot] coordinates {};
\end{axis}
\end{tikzpicture}
  \vspace{-10pt}
  \caption{RAN to dApp latency in the
           L1 SM, for \gls{asn1} and \gls{json} indication encodings in \gls{dl} and \gls{ul}.}
  \label{fig:latency_box}
  \vspace{-.5cm}
\end{figure}

As shown in Figure~\ref{fig:latency_box}, across all four configurations (ASN.1 DL/UL and JSON DL/UL), the median latency remains between approximately 115 and 120\,\si{\micro\second}, with an interquartile range of only about 10\,\si{\micro\second}. These values remain well below the sub-millisecond latency budget required by real-time dApps under heavy traffic. Latency is largely insensitive to traffic direction and encoding, with nearly identical results across configurations.

\begin{figure}[t]
  \centering
  \setlength{\fwidth}{\columnwidth}\setlength{\fheight}{0.42\columnwidth}
\definecolor{darkgray176}{RGB}{176,176,176}
\definecolor{cencode}{RGB}{127,186,176}
\begin{tikzpicture}
\begin{axis}[
width=\fwidth, height=\fheight,
font=\scriptsize,
xlabel={Time [\si{\micro\second}]},
xmin=0, xmax=115,
ymin=-0.24, ymax=3.84,
ytick={0.4,1.4,2.4,3.4},
xmajorgrids, ymajorgrids,
grid style={darkgray176},
yticklabels={
  {Asn1 DL},
  {Asn1 UL},
  {JSON DL},
  {JSON UL}
},
legend style={font=\scriptsize, fill opacity=0.85, draw opacity=1, text opacity=1, draw=darkgray176, at={(0.5,1.03)}, anchor=south, legend columns=-1},
tick label style={font=\scriptsize},
label style={font=\scriptsize},
xlabel shift=-3pt,
]
\path [fill=yellow!20,postaction={pattern=vertical lines, pattern color=black}]
(axis cs:0.00,0.00) --(axis cs:0.00,0.80) --(axis cs:69.02,0.80) --(axis cs:69.02,0.00) --cycle;
\path [fill=cencode,postaction={pattern=crosshatch, pattern color=black}]
(axis cs:69.02,0.00) --(axis cs:69.02,0.80) --(axis cs:70.25,0.80) --(axis cs:70.25,0.00) --cycle;
\path [fill=green!10,postaction={pattern=north west lines, pattern color=black}]
(axis cs:70.25,0.00) --(axis cs:70.25,0.80) --(axis cs:81.39,0.80) --(axis cs:81.39,0.00) --cycle;
\path [fill=orange!15,postaction={pattern=north east lines, pattern color=black}]
(axis cs:81.39,0.00) --(axis cs:81.39,0.80) --(axis cs:106.19,0.80) --(axis cs:106.19,0.00) --cycle;
\path [fill=yellow!20,postaction={pattern=vertical lines, pattern color=black}]
(axis cs:0.00,1.00) --(axis cs:0.00,1.80) --(axis cs:69.69,1.80) --(axis cs:69.69,1.00) --cycle;
\path [fill=cencode,postaction={pattern=crosshatch, pattern color=black}]
(axis cs:69.69,1.00) --(axis cs:69.69,1.80) --(axis cs:70.89,1.80) --(axis cs:70.89,1.00) --cycle;
\path [fill=green!10,postaction={pattern=north west lines, pattern color=black}]
(axis cs:70.89,1.00) --(axis cs:70.89,1.80) --(axis cs:81.72,1.80) --(axis cs:81.72,1.00) --cycle;
\path [fill=orange!15,postaction={pattern=north east lines, pattern color=black}]
(axis cs:81.72,1.00) --(axis cs:81.72,1.80) --(axis cs:105.73,1.80) --(axis cs:105.73,1.00) --cycle;
\path [fill=yellow!20,postaction={pattern=vertical lines, pattern color=black}]
(axis cs:0.00,2.00) --(axis cs:0.00,2.80) --(axis cs:68.87,2.80) --(axis cs:68.87,2.00) --cycle;
\path [fill=cencode,postaction={pattern=crosshatch, pattern color=black}]
(axis cs:68.87,2.00) --(axis cs:68.87,2.80) --(axis cs:72.61,2.80) --(axis cs:72.61,2.00) --cycle;
\path [fill=green!10,postaction={pattern=north west lines, pattern color=black}]
(axis cs:72.61,2.00) --(axis cs:72.61,2.80) --(axis cs:83.64,2.80) --(axis cs:83.64,2.00) --cycle;
\path [fill=orange!15,postaction={pattern=north east lines, pattern color=black}]
(axis cs:83.64,2.00) --(axis cs:83.64,2.80) --(axis cs:108.50,2.80) --(axis cs:108.50,2.00) --cycle;
\path [fill=yellow!20,postaction={pattern=vertical lines, pattern color=black}]
(axis cs:0.00,3.00) --(axis cs:0.00,3.80) --(axis cs:69.70,3.80) --(axis cs:69.70,3.00) --cycle;
\path [fill=cencode,postaction={pattern=crosshatch, pattern color=black}]
(axis cs:69.70,3.00) --(axis cs:69.70,3.80) --(axis cs:73.24,3.80) --(axis cs:73.24,3.00) --cycle;
\path [fill=green!10,postaction={pattern=north west lines, pattern color=black}]
(axis cs:73.24,3.00) --(axis cs:73.24,3.80) --(axis cs:83.96,3.80) --(axis cs:83.96,3.00) --cycle;
\path [fill=orange!15,postaction={pattern=north east lines, pattern color=black}]
(axis cs:83.96,3.00) --(axis cs:83.96,3.80) --(axis cs:108.04,3.80) --(axis cs:108.04,3.00) --cycle;
\addlegendimage{fill=yellow!20,postaction={pattern=vertical lines, pattern color=black},area legend}\addlegendentry{Grid + SHM write}
\addlegendimage{fill=cencode,postaction={pattern=crosshatch, pattern color=black},area legend}\addlegendentry{E3SM encode}
\addlegendimage{fill=green!10,postaction={pattern=north west lines, pattern color=black},area legend}\addlegendentry{Emit}
\addlegendimage{fill=orange!15,postaction={pattern=north east lines, pattern color=black},area legend}\addlegendentry{Publish (libE3)}
\end{axis}
\end{tikzpicture}
  \vspace{-10pt}
   \vspace{-12pt}
  \caption{\emph{E3Controller} L1 SM pipeline latency for \gls{asn1}/\gls{json} encodings in \gls{dl}/\gls{ul}.}
  \label{fig:latency_timeline}
  \vspace{-.3cm}
\end{figure}
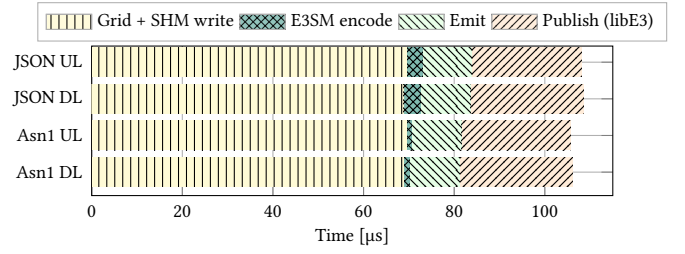

To identify the main contributors, Figure~\ref{fig:latency_timeline} decomposes the data path in Figure~\ref{fig:flow} into the median duration of four stages. The grid transfer (\gls{ran} to codelet) and shared-memory write, which convert the resource grid and write it to buffers, are the only stages handling the I/Q payload and dominate at approximately $69$\,\si{\micro\second}, primarily due to memory bandwidth and data conversion rather than codelet logic. The \gls{e3sm} encode and the \gls{sm} emit each add only a few microseconds, while the publisher, which performs \gls{e3ap} encoding and transmission through \texttt{libE3}, adds about $24$\,\si{\micro\second}. The intermediate codelet message poller (\emph{JbpfDispatcher}) stays sub-microsecond and is omitted. The four stages account for approximately $106$\,\si{\micro\second}.  The residual latency relative to Figure~\ref{fig:latency_box} is attributable to queueing in \texttt{libE3} prior to indication dequeueing.

\begin{figure}[t]
  \centering
  \setlength{\fwidth}{\columnwidth}\setlength{\fheight}{0.35\columnwidth}
\definecolor{darkgray176}{RGB}{176,176,176}
\definecolor{casn}{RGB}{127,186,176}
\definecolor{cjson}{RGB}{181,140,198}
\begin{tikzpicture}
\begin{axis}[
width=\fwidth,
height=\fheight,
font=\scriptsize,
legend cell align={left},
legend style={font=\scriptsize, fill opacity=0.85, draw opacity=1, text opacity=1, draw=darkgray176, at={(0.5,1.03)}, anchor=south, legend columns=-1},
tick align=outside,
tick pos=left,
grid style={darkgray176},
ymajorgrids,
ylabel style={font=\scriptsize},
tick label style={font=\scriptsize},
ylabel shift=-3pt,
ylabel={Encoding latency [\si{\micro\second}]},
boxplot/draw direction=y,
boxplot/box extend=0.55,
boxplot/every median/.style={black, line width=0.9pt},
xtick={1.5,4.5},
xticklabels={\gls{dl},\gls{ul}},
xmin=0.3, xmax=5.7,
ymin=0.2, ymax=6.1,
]
\addlegendimage{area legend, fill=casn, draw=black, postaction={pattern=north east lines, pattern color=black}}
\addlegendentry{Asn1}
\addlegendimage{area legend, fill=cjson, draw=black}
\addlegendentry{JSON}
\addplot[boxplot prepared={draw position=1, lower whisker=0.59, lower quartile=1.01, median=1.23, upper quartile=1.51, upper whisker=2.26},
  fill=casn, draw=black, line width=0.5pt, forget plot] coordinates {};
\addplot[boxplot prepared={draw position=1, lower whisker=0.59, lower quartile=1.01, median=1.23, upper quartile=1.51, upper whisker=2.26},
  pattern=north east lines, pattern color=black, draw=black, line width=0.5pt, forget plot] coordinates {};
\addplot[boxplot prepared={draw position=2, lower whisker=2.18, lower quartile=3.37, median=3.74, upper quartile=4.26, upper whisker=5.59},
  fill=cjson, draw=black, line width=0.5pt, forget plot] coordinates {};
\addplot[boxplot prepared={draw position=4, lower whisker=0.52, lower quartile=0.99, median=1.20, upper quartile=1.43, upper whisker=2.07},
  fill=casn, draw=black, line width=0.5pt, forget plot] coordinates {};
\addplot[boxplot prepared={draw position=4, lower whisker=0.52, lower quartile=0.99, median=1.20, upper quartile=1.43, upper whisker=2.07},
  pattern=north east lines, pattern color=black, draw=black, line width=0.5pt, forget plot] coordinates {};
\addplot[boxplot prepared={draw position=5, lower whisker=2.23, lower quartile=3.22, median=3.54, upper quartile=3.91, upper whisker=4.93},
  fill=cjson, draw=black, line width=0.5pt, forget plot] coordinates {};
\end{axis}
\end{tikzpicture}
  \vspace{-10pt}
  \caption{\gls{e3sm} encoding latency in the L1 SM for \gls{asn1} and \gls{json} indication encodings in \gls{dl} and \gls{ul}.}
  \label{fig:latency_encode}
  \vspace{-.5cm}
\end{figure}
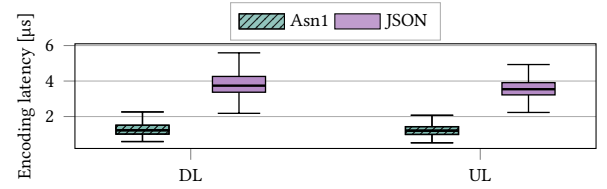

Figure~\ref{fig:latency_encode} isolates the \gls{e3sm} encoding stage, the only component that differs between \gls{asn1} and \gls{json}. \gls{json} encodes in about $3.6$\,\si{\micro\second} against $1.2$\,\si{\micro\second} for \gls{asn1}, roughly three times slower, yet remains negligible relative to the overall pipeline. Thus, encoding has no measurable impact on throughput (Figure~\ref{fig:cdf}) or energy (Table~\ref{tab:energy}), which are dominated by resource-grid transfer.

\section{Conclusion and Future Work}

This paper presented the first integration of the E3 interface with OCUDU, enabling \gls{e3ap}-compliant dApps to access real-time \ran data through a well-defined interface. The proposed architecture combines lightweight in-\ran instrumentation with an external \texttt{E3Controller}, preserving the modularity of the \ran implementation while exposing PHY- and fronthaul-level information to external applications. Experimental evaluation demonstrated that the platform delivers high-rate I/Q streams to latency-sensitive dApps without measurable impact on \ran throughput, while maintaining end-to-end latencies well within the sub-millisecond budget required by real-time control applications.

Looking ahead, we envision the \texttt{E3Controller} as a general execution platform for programmable \ran services rather than solely a data-export mechanism. Enabling dApps to dynamically inject logic into selected \ran functions would allow runtime extension of scheduling, PHY processing, and other protocol-stack components without modifying or redeploying the underlying \ran software. Such a capability would transform dApps into active extensions of the \ran processing pipeline, enabling ultra-low-latency optimization, sensing, and AI-driven applications.

\begin{acks}
This work was supported by OUSW(R\&E) through Army Research Laboratory
Cooperative Agreement Number W911NF-24-2-0065. The views and conclusions
contained in this document are those of the authors and should not be
interpreted as representing the official policies, either expressed or implied,
of the Army Research Laboratory or the U.S. Government. The U.S. Government is
authorized to reproduce and distribute reprints for Government purposes
notwithstanding any copyright notation herein.

\end{acks}
\vspace{-.5cm}

\bibliographystyle{ACM-Reference-Format}
\bibliography{reference}

\end{document}